\documentclass[bibyear]{aa}

\usepackage{graphics}
\usepackage{graphicx}
\usepackage[varg]{txfonts}
\usepackage{psfig}
\usepackage{float}
\usepackage{psfrag}
\usepackage{amsmath}
\usepackage{amssymb}
\usepackage{bbm}
\usepackage{ulem}
\usepackage{threeparttable}
\usepackage{array}
\usepackage{multirow}
\usepackage{enumerate}
\usepackage[usenames]{xcolor}

\usepackage{natbib,twoopt}
\definecolor{darkgreen}{rgb}{0,0.5,0}
\usepackage[breaklinks=true,colorlinks=true,linkcolor=blue,citecolor=darkgreen,linkbordercolor=blue,citebordercolor=magenta,urlbordercolor=green]{hyperref}
\bibpunct{(}{)}{;}{a}{}{,}

\newcommand{\plk}{\textit{Planck }}

\newcommand{\asec}{^{\prime\prime}}
\newcommand{\mach}{{\cal M}}
\newcommand{\sausage}{\textit{Sausage}~}
\newcommand{\toothbrush}{\textit{Toothbrush}~}

\newcommand{\rev}{\normalfont}
\newcommand{\mathcorrect}{\normalfont}
\newcommand{\last}{\normalfont}
\newcommand{\syn}{{\normalfont{synchrotron} }}

\begin{document}

\title{The impact of the SZ effect on cm-wavelength (1--30 GHz) \\ observation of galaxy cluster radio relics}
\author{Kaustuv Basu\inst{1},~ Franco Vazza\inst{2},~ Jens Erler\inst{1}~\& Martin Sommer\inst{1}} 
\institute{
     	Argelander Institut f\"ur Astronomie, Universit\"at Bonn, Auf dem H\"ugel 71, 53121 Bonn, Germany\\
	\email{kbasu@astro.uni-bonn.de}
	\and
	Hamburger Sternwarte, Gojenbergsweg 112, 21029 Hamburg, Germany \\
	}

%%%%%%%%%%%%%%%%%%%%%%%%%%%%%%%%%%%%%%%%%%%%%%%%%%%%%%%%%%%%%%%%%%%%%%%%%%%%%%%%%%%%%%%
\abstract{
Radio relics in galaxy clusters are believed to be associated with powerful shock fronts that originate during cluster mergers, and are a testbed for the acceleration of relativistic particles in the intracluster medium. Recently, radio relic observations have pushed into the cm-wavelength domain ($1-30$ GHz) where a break from the standard synchrotron power-law spectrum has been found, most noticeably in the famous `Sausage' relic. Such spectral steepening is seen as an evidence for non-standard relic models, such as ones requiring seed electron population with a break in their energy spectrum. 
In this paper, however, we point to an important effect that has been ignored or considered insignificant while interpreting these new high-frequency radio data, namely the contamination due to the Sunyaev-Zel'dovich (SZ) effect that changes the observed \syn flux. Even though the radio relics reside in the cluster outskirts, the shock-driven pressure boost increases the SZ signal locally by roughly an order of magnitude. The resulting flux contamination for some well-known relics are non-negligible already at 10 GHz, and at 30 GHz the observed \syn fluxes can be diminished by a factor of several from their true values. At higher redshift the contamination gets stronger due to the redshift independence of the SZ effect. Interferometric observations are not immune to this contamination, since the change in the SZ signal occurs roughly at the same length scale as the \syn emission, although there the flux loss is less severe than single-dish observations. 
Besides presenting this warning to observers, we suggest that the negative contribution from the SZ effect can be regarded as one of the best evidence for the physical association between radio relics and shock waves. We present a simple analytical approximation for the \rev{synchrotron}-to-SZ flux ratio, based on a theoretical radio relic model that connects the non-thermal emission to the thermal gas properties, and show that by measuring this ratio one can potentially estimate the relic magnetic fields or the particle acceleration efficiency.
}
%%%%%%%%%%%%%%%%%%%%%%%%%%%%%%%%%%%%%%%%%%%%%%%%%%%%%%%%%%%%%%%%%%%%%%%%%%%%%%%%%%%%%%%

 % -- radio continuum: general}

\date{Received 12 November 2015 / Accepted 1 April 2016}

\titlerunning{SZ contamination in radio relics}
\authorrunning{K. Basu et al.}
\maketitle

%%%%%%%%%%%%%%%%%%%%%%%%%%%%%%%%%%%%%%%%%%%%%%%%%%%%%%%%%%%%%%%%%%%%%%%%%%%%%%%%%%%%%%%
\section{Introduction}
%%%%%%%%%%%%%%%%%%%%%%%%%%%%%%%%%%%%%%%%%%%%%%%%%%%%%%%%%%%%%%%%%%%%%%%%%%%%%%%%%%%%%%%

Along with their cosmological relevance, galaxy clusters are giant laboratories for studying many interesting astrophysical processes. It is often the case that different astrophysical phenomena, which are observed in completely different regimes of the electromagnetic spectrum, are connected by the same underlying cause. A prominent example of such a case is cluster merger shocks. Shocks cause a jump in the density and temperature  of the thermal intra-cluster medium (ICM), and  are believed to produce non-thermal cosmic ray electrons (CRe) through the first-order Fermi acceleration. The former effect can be observed in the soft X-ray band or in the millimeter/sub-millimeter wavebands through the Sunyaev-Zel'dovich effect (thermal SZ effect, tSZ effect, or simply SZ for short; \citealt{SZ72}, \citeyear{SZ80}), whereas the latter is visible in the radio wavebands through synchrotron emission or in the hard X-rays through inverse Compton emission. This paper aims to explore some observational consequence when the radio synchrotron signal and the SZ effect overlap each other at cm-wavelengths (1-30 GHz).

Clusters grow through mergers and these mergers drive shock waves through the ICM \citep[e.g.,][]{Sar02}. In cluster simulations shocks are ubiquitous (e.g. \citealt{Nuz12}, \citealt{Skill13}, \citealt{Hong14}) but their direct observation is still limited by small number statistics as detecting ICM shocks are hampered by their low surface brightness and projection effects, particularly in the cluster outskirts where the thermal signature is hard to detect in SZ or X-rays \rev{(e.g., \citealt{Aka13}, \citealt{PComa}, \citealt{Er15})}. 
Radio relics are elongated, diffuse synchrotron sources found in the cluster outskirts, have length scales of the order of $\sim 1$ Mpc and are not associated with any AGN outflow (e.g. review by \citealt{Fer12}). The common wisdom is that these radio relics mark the location of powerful shock fronts generated by cluster mergers, first  postulated by \citet{Enss98}. The mechanism behind the acceleration of GeV energy electrons at the shock fronts is believed to be diffusive shock acceleration (DSA; \citealt{Bland87}), and have been the subject of extensive studies both analytically and through numerical simulations (e.g. \citealt{HB07}, \citealt{Pfrom08}, \citealt{Nuz12}, \citealt{Kang12}, \citealt{Skill13}, \citealt{Pinz13}, \citealt{Vaz14}, \citealt{Vaz15}). 
However, the problem with the DSA-based scenario is the low Mach number of merger shocks inside a cluster volume ($\mach \lesssim 4$; \citealt{Ryu03}, \citealt{Pfrom06}), which would make the acceleration efficiency prohibitively low. To overcome this problem special modification to the ambient Maxwell-J\"uttner distribution of thermal electrons have been proposed, e.g. in the form of a $\kappa$-distribution to add a suprathermal tail \citep[e.g.,][]{Kang14}, or by postulating a pre-existing population of CRe from past AGN activities at the cluster core \citep[e.g.,][]{Pinz13}. On the other hand, recent particle-in-cell simulations have suggested alternative mechanisms based on shock drift acceleration (SDA) that can efficiently inject thermal electrons into DSA-like cycle, thus eliminating any need for additional non-thermal population beforehand (\citealt{Mats11}, \citealt{Guo14a}, \citeyear{Guo14b}). Thus it is of utmost importance to find observational signatures of the underlying CRe energy distribution that causes radio relics, to find out whether a seed population of CRe exists or not prior to the shock passage.

One way of knowing the particle injection spectrum and the associated magnetic field in the relics is to measure the relic fluxes out to high frequencies, looking for spectral break in the relic spectrum that can indicate deviation from a power-law for the seed CRe population. Recently, some spectacular examples of a spectral-break in relic spectrum have been published in the literature, in particular for the famous \sausage relic in the cluster CIZA J2242.8$+$5301 (\citealt{Str14}, \citeyear{Str15}). Based on the radio interferometric data obtained at 16 GHz, \citet{Str14} first found a pronounced steepening of the integrated relic spectrum at high frequencies, which is inconsistent with the extrapolation of the low-frequency spectrum at $12\sigma$. 
%This spectral curvature was then used as an argument for an ageing electron population or strong magnetic field downstream.  
Recent results for the  \sausage and \toothbrush relics published by \citet{Str15} extends the spectral measurements out to 30 GHz, with consistent trend for spectral steepening. 
\rev{Such steepening at high frequencies is not expected according to the standard DSA-based acceleration scenario, since there is a balance between particle injection and ageing, hence these recent observations have been used as an argument for the presence of aged fossil electrons or strong magnetic field downstream (\citealt{Str15}).} 
Another famous relic that was observed out to 10 GHz is in the cluster A2256 \citep{Tra15}, also showing mild evidence of spectral steepening. 

In light of these new results, a careful investigation of the amplitude of the SZ signal at radio relic location is necessary,  a problem which has not been investigated in detail so far. The current paper tries to address this issue comprehensively, with a semi-analytical approach.  
For frequencies below 220 GHz the SZ effect appears as a negative signal against the CMB background. At the cm-wavebands of interest ($1-30$ GHz) the \rev{synchrotron} power drops rapidly due to its steep spectrum, whereas the SZ effect increases in strength as $\nu^2$; furthermore, at the location of relics the SZ signal gets boosted by the shock roughly proportional to the Mach number squared. Since the SZ signal boost occurs within the same spatial scales responsible for the radio relic emission, the decrement is \rev{not completely filtered out} by  interferometric observations.
We show in this paper that for many relics the flux modification due to the SZ effect is non-negligible: \rev{it can reduce} the observed \rev{synchrotron} flux by $\sim 10-50\%$ at 15 GHz, and up to a factor of several at 30 GHz. These estimates are mostly for radio relic clusters at $z\lesssim 0.2$, and the contamination increases rapidly with redshift. 
In fact, far from being a simple ``nuisance'' for observations, the SZ decrement at this frequency range can offer probably the best evidence for the widely accepted relic-shock connection. Our modeling shows that, within a factor $\sim 2$ or so, the predicted SZ decrement is consistent with the reported high frequency steepening of relics.

The layout of the paper is as following. In Section \ref{sec:basics} we briefly describe the theory behind the SZ effect and the \rev{synchrotron} emission from radio relics, in particular how these signals are expected to behave across a shock front. We introduce an empirical ``toy model'' for the \rev{synchrotron emissivity} profile, and illustrate the flux contamination issue rev{realistically} with simulations for interferometric observation.  
Section \ref{sec:allresults} presents the results: first some general characteristics of the SZ flux contamination, e.g. how the signals will change with cluster mass, redshift and the shock Mach number, and then individual estimates for the flux contamination in some well known radio relics. In Section \ref{sec:disc} we provide an analytical approximation for the radio \rev{synchrotron-to-SZ} flux ratio, and discuss how its measurement can help to refine the theory of particle acceleration for radio relic origin or constrain the relic magnetic fields. We summarize our work in Section \ref{sec:con}, and conclude with some general warnings for observers and suggestions for future theoretical work. 
%some warning for observers to interpret their GHz-frequency radio relic spectrum. 
For computing the distances and measured fluxes, we assume a standard $\Lambda$CDM cosmology throughout the paper: $H_0$= 71 km s$^{-1}$ Mpc$^{-1}$, $\Omega_m=0.27$ and $\Omega_{\Lambda}=0.73$. The redshift-dependent Hubble ratio is defined as $E(z) \equiv H(z)/H_0 = \sqrt{(1+z)^3\Omega_m + \Omega_{\Lambda}}$.

%%%%%%%%%%%%%%%%%%%%%%%%%%%%%%%%%%%%%%%%%%%%%%%%%%%%%%%%%%%%%%%%%%%%%%%%%%%%%%%%%%%%%%%
\section{Theoretical and observational basics}
\label{sec:basics}
%%%%%%%%%%%%%%%%%%%%%%%%%%%%%%%%%%%%%%%%%%%%%%%%%%%%%%%%%%%%%%%%%%%%%%%%%%%%%%%%%%%%%%%

We describe the basics of the Sunyaev-Zel'dovich (SZ) effect, and explain how the SZ and \syn signals are boosted by the passage of a shock.  To model the \syn flux realistically, we construct an empirical model and compare with published relic emission profiles. We also outline  the main differences between interferometric and single-dish imaging techniques, and present simulation results for interferometric observation of radio relics in presence of a non-negligible SZ signal.

\subsection{``Negative flux'' from the SZ effect}
\label{sec:sz}

The SZ effect at cm-wavelengths acts as a contaminant to the \rev{synchrotron} flux, making the observed flux less than its true value. This is because at the cm- to mm-wavelengths the SZ signal appears as a decrement in the background CMB radiation intensity. Since all flux measurements are done against the backdrop of the CMB radiation, which defines the ``zero level'' of flux, for all practicality we can consider the SZ signal as a ``true'' negative signal as opposed to the positive synchrotron flux (for example). 
%In practice, the observational method will determine how the negative and positive signals are sampled, which we discuss in Sec. \ref{sec:interferometry} in the context of single-dish and interferometric observation of radio relics. 
\rev{In practice, interferometric or single-dish observation sample the distribution of positive and negative fluxes differently, resulting in very different total power images. This we discuss in Sec. \ref{sec:interferometry}. }

The change in the CMB temperature due to the thermal SZ effect is given by the well-known formula
\begin{equation}
\Delta T / T_{\mathrm{CMB}} = f(x)~ y,
\end{equation}
where $x$ is the dimensionless frequency, $y$ is the Comptonization parameter, $T_{\mathrm{CMB}} = 2.726$ K, and the frequency spectrum of the temperature variation has the form
\begin{equation}
f(x) = \left(x\dfrac{e^x+1}{e^x-1} - 4\right)\left[1 + \delta_{\mathrm{rSZ}}(x, T_e)\right] 
\end{equation}
The second term in parenthesis is the relativistic correction to the SZ effect, and even though the passage of a merger shock is expected to heat up the ICM to $\sim 10$ keV or more right behind the shock, this relativistic term can be safely ignored in the frequency range of interest ($\lesssim 30$ GHz). In the Rayleigh-Jeans limit ($x \lesssim 0.5$) we thus get $f(x) \approx -2$, and the change in brightness temperature is {\it negative}. Converting $\Delta T_{\mathrm{CMB}}$ to Rayleigh-Jeans temperature by multiplying with $x^2e^x/(e^x-1)^2$, one can obtain the following simplified expression for the measured negative flux density in a given instrument beam (e.g. \citealt{Birk99}):
\begin{equation}
\left(\dfrac{S_{\nu}}{\mathrm{mJy/beam}}\right) = \dfrac{1}{340} \left(\dfrac{\Delta T_{\mathrm{RJ}}}{\mathrm{mK}}\right) 
	\left(\dfrac{\nu}{\mathrm{GHz}}\right)^2 \left(\dfrac{\Omega_{\mathrm{beam}}}{\mathrm{arcmin}^2}\right). 
\label{eq:szflux}
\end{equation}
$\Omega_{\mathrm{beam}}$ is the beam solid angle, relating to the half-power beam width as approximately $\Omega_{\mathrm{beam}} = 1.13~ \theta_{\mathrm{FWHM}}^2$. 
The factor 1.13 is dropped when we express the SZ flux density independent of telescope beams, in units of mJy/arcmin$^{2}$.

The Comptonization parameter, $y$, is obtained through the line-of-sight integral of the cluster pressure distribution
\begin{equation}
y = \dfrac{\sigma_{\mathrm{T}}}{m_e c^2}~ \int_{\mathrm{l.o.s.}} P_e(r)~ \mathrm{d}l 
\label{eq:y}
\end{equation}
For the spherical pressure distribution, we shall assume the generalized NFW, or GNFW, pressure model \citep{Nag07}, with the mass dependence and parametrization as given by \citet{Ar10}:
\begin{equation}
P_e(r) = 1.65\times 10^{-3} E(z)^{8/3} \left[\frac{M_{500}}{3\times 10^{14} M_{\odot}}\right]^{0.79} \mathbbm{p}\left(\frac{r c_{500}}{r_{500}}\right)	~ \mathrm{keV cm}^{-3}
\label{eq:gnfw}
\end{equation}
The term $\mathbbm{p}(r c_{500}/r_{500})$ gives the ``universal'' shape of the cluster pressure profile, with various slope parameters and the gas concentration parameter $c_{500}$. In the cluster outskirts where radio relics are located, the pressure falls-off rapidly, like $P(r) \propto r^{-5.5}$ or similar, hence a low amplitude of the SZ signal  is expected. 
There is about 30\% scatter in the cluster pressure amplitude (\citealt{Ar10}, \citealt{Ppress13}, \citealt{Say13}), but this is within our accuracy range for predicting  the SZ contamination. In Section \ref{sec:indiv} we further re-normalize this GNFW pressure model results with individual cluster's SZ measurement whenever possible, to make our flux contamination predictions more accurate.

The above is the standard formulation for the SZ signal in a cluster, which produces a ``bowl'' of negative flux, whose amplitude near the cluster virial radius is orders of magnitude lower than the peak decrement value at the center. But we are dealing with a shock, and the shock will boost the SZ signal locally. This can be computed from the standard Rankine-Hugoniot condition of pressure jump at the location of a shock: 
\begin{equation}
\dfrac{P_\mathrm{d}}{P_\mathrm{u}} = \dfrac{2\gamma \mach^2 - (\gamma-1)}{(\gamma+1)}.
\label{eq:Pratio}
\end{equation}
$\mach$ is the sonic Mach number of the shock, $\gamma = 5/3$ is the adiabatic index, and we define $P_\mathrm{d}$ and $P_\mathrm{u}$ are the instantaneous pressure in the downstream (post-shock) and upstream (pre-shock) regions. We see that the pressure jump scales roughly as $\sim \mach^2$, hence the boost in the SZ signal is non-negligible. Roughly speaking, this creates a ``step-function'' like increase in the SZ flux decrement at the location of the radio relic on top of the GNFW ``bowl''. 
%causing a significant flux contamination.
The sharp change in the SZ signal is also important in the context of interferometric observation, as discussed in Section \ref{sec:interferometry}.

\begin{figure}[h]
\centering
\includegraphics[width=0.9\columnwidth, height=5cm]{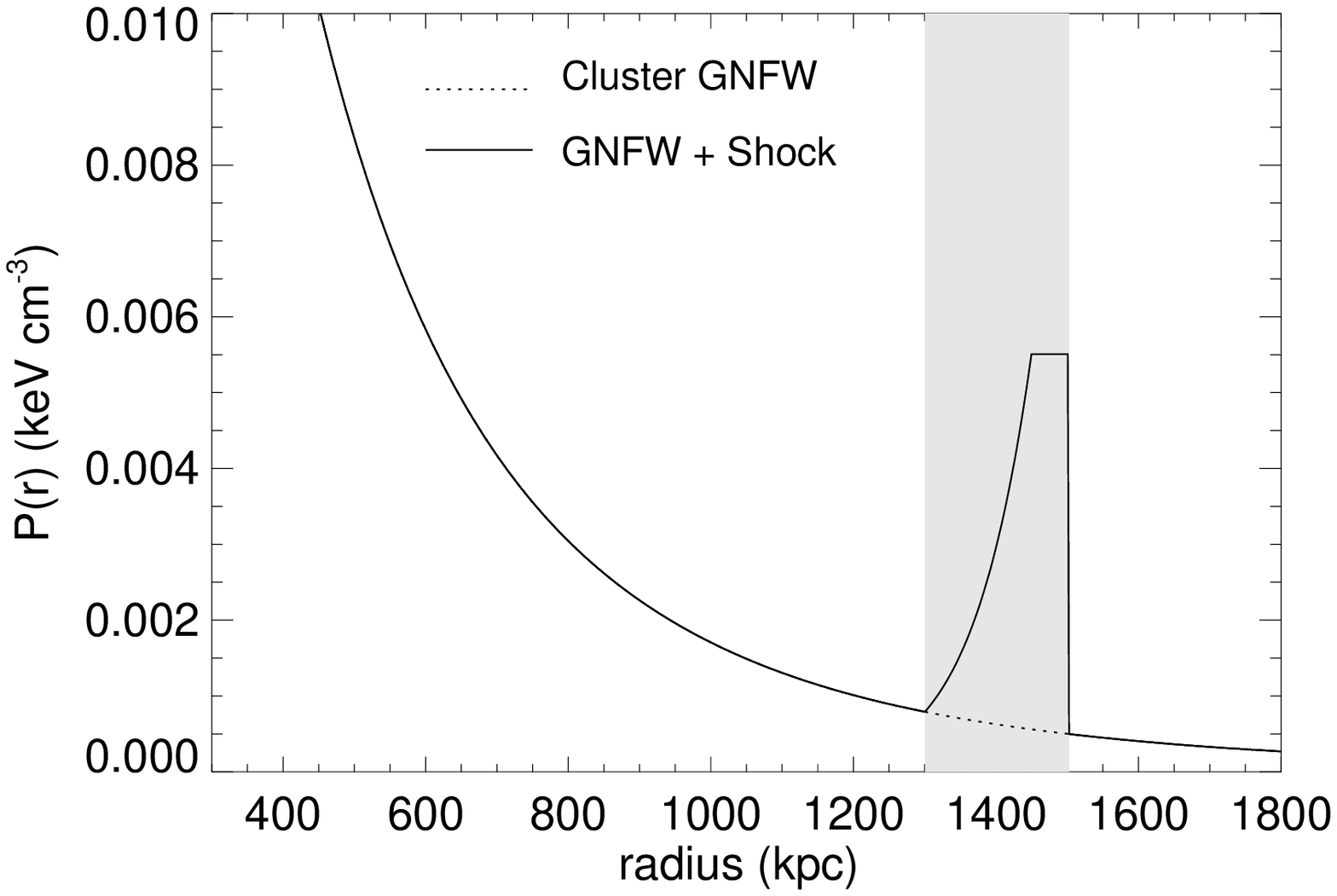}
\includegraphics[width=0.9\columnwidth, height=5.cm]{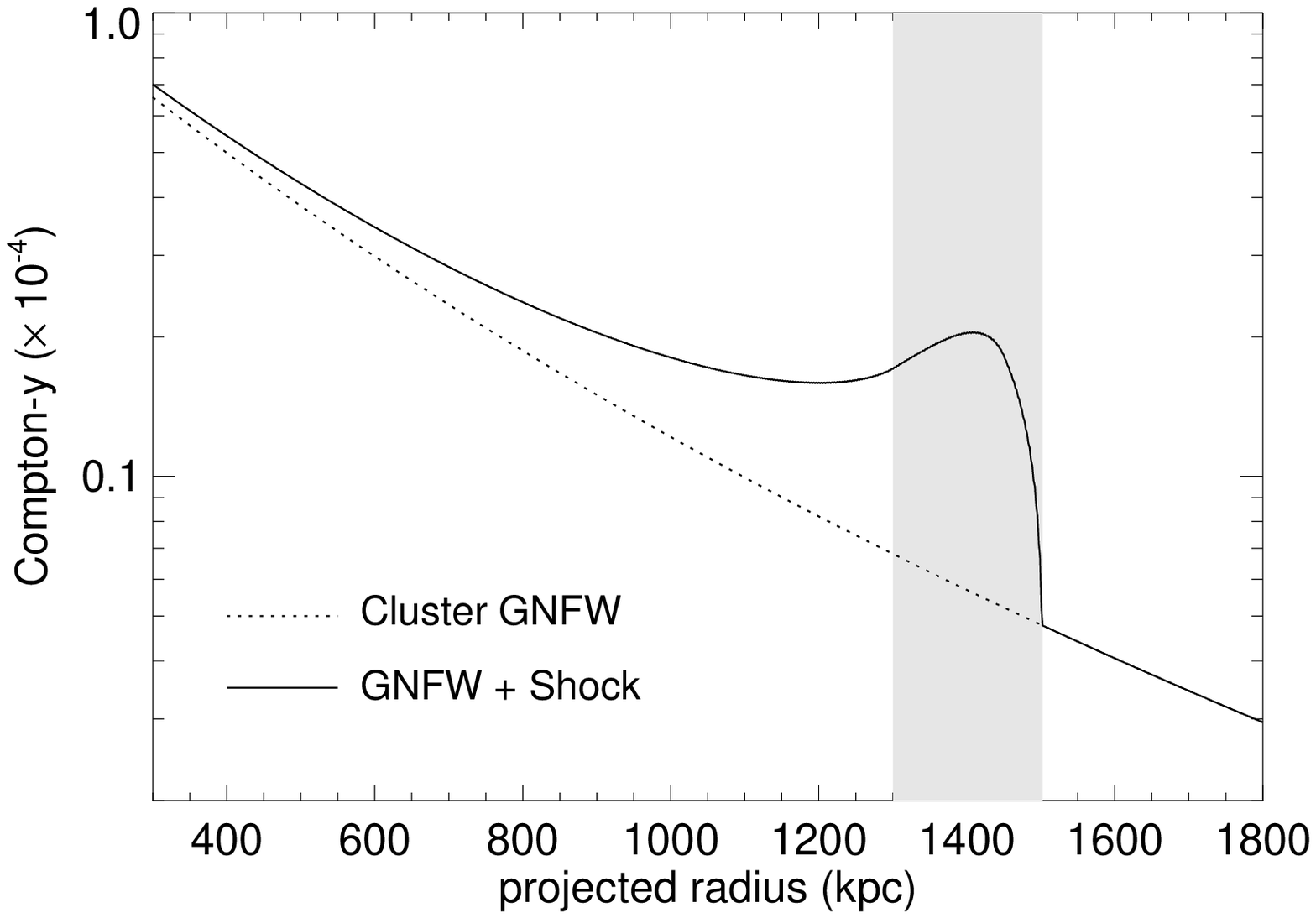}
\includegraphics[width=0.9\columnwidth, height=5cm]{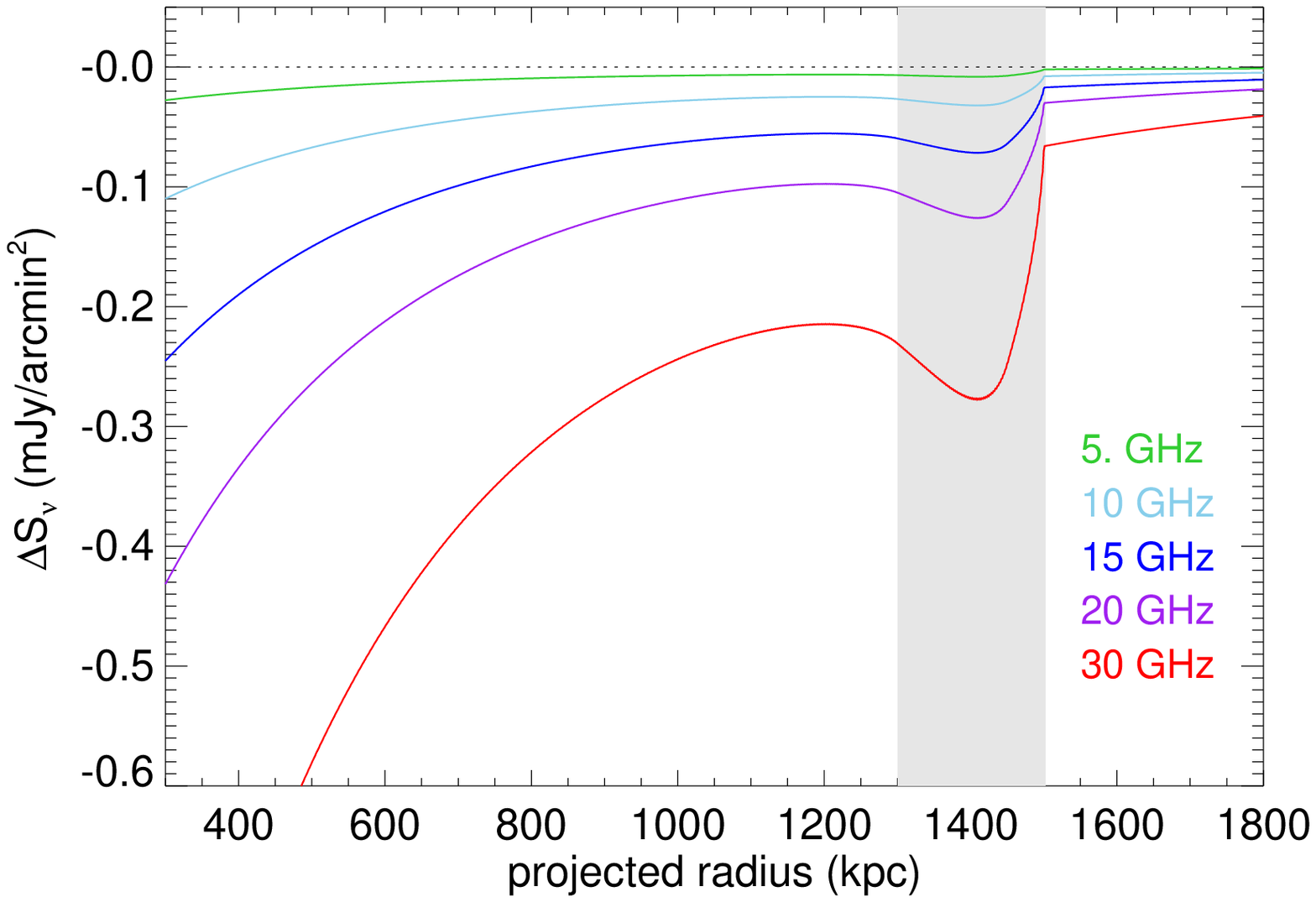}
\caption{Illustration of SZ flux decrement at a merger shock, for a spherical shock model at 1.5 Mpc distance from the cluster center, in a $M_{500}=8\times 10^{14}$ M$_{\odot}$ cluster at $z=0.2$, with shock Mach number $\mach=3$. 
The gray vertical bands mark the approximate pressure-boosted region in radial distance. 
{\it Top panel:} The radial distribution of pressure, demonstrating the localized boost in pressure by roughly factor $\sim 10$.  The underlying dotted line is the ambient GNFW pressure model proportional to the mass of the cluster. 
{\it Middle panel:} The projected Compton-$y$ parameter, solid line showing the total value.  
{\it Bottom panel:} The resulting SZ flux densities, in units of mJy/arcmin$^2$, as would be observed at different frequencies.} 
\label{fig:szshock}
\end{figure}

To make a realistic model for the SZ flux variation, we need to assume a shock geometry as well as some cooling length behind the shock. For the former, we shall assume a standard spherical shock geometry. This is clearly a simplification, as simulations show that although shock fronts radiate outwards in all direction after a major merger, its geometry is not actually spherical due to the inhomogeneous structure of the ICM  (e.g. \citealt{Mini00}). 
However, for our purposes it suffices that the SZ signal boost is localized within the same geometrical boundary as marked by the relics, because the flux contamination will be determined primarily from the projection of pressure from a narrow annular section of the shock where the relic resides. 
The second point is more complicated, namely the width of the shock-boosted region. Lacking any sophisticated simulations for gas cooling, we follow the examples from shock-tube geometry (e.g. \citealt{Mini07}, \citealt{Vaz12}), simply assuming that the boosted pressure stays constants for a small distance (which we can call the shock width) after which it drops due to cooling with a power-law and merges into the ambient pressure. This setting is shown in Fig. \ref{fig:szshock}  top panel. We take a small shock-width of 50 kpc, in line with the most resolved available radio observations (e.g. \citealt{vanWe10}). We ignore any dependence of this width with the shock Mach number. However, we can state that the SZ contamination on radio relic fluxes is not very sensitive to the precise model of the downstream pressure: this is because the radio emission is mostly concentrated just behind the shock front (Section \ref{sec:radio} below), where the pressure closely follows the boosted value according to equation \ref{eq:Pratio}.

We numerically project this 3D pressure model to get the Compton-$y$ parameter and the expected flux decrement at any given frequency. The results are shown in Fig. \ref{fig:szshock} for a fiducial shock model, for a cluster with similar mass and redshift as in the \sausage relic host cluster. The projected distances  are expressed in kpc for ease of comparison with the pressure profile. We see that due to projection the effect of the SZ signal boost extends far inwards relative to the shock front, but its amplitude exhibits a local maximum right at the location of the shock width after rising quickly from the ambient SZ signal. 
%is roughly constant after quickly rising to its maximum behind the shock front. 
In other words, the SZ signal {\it variation} is strongest in a relatively narrow region, of the order $\sim 100$ kpc, where the radio relic is also at its brightest. 
\rev{In our SZ modeling we are neglecting the effect of non-equilibrium between the electrons and ions in the post-shock region. From a simple estimate based on Coulomb interaction time one can expect electron thermalization to happen over a length-scale of $\sim 50$ kpc for a $\mach \sim 3$ shock, but this represents an upper limit of the effect \citep{Rud09}, because it neglects energy exchange mediated by plasma instabilities which can shorten this equilibration timescale significantly (e.g. \citealt{Byk08}). 
% because the thermalization time-scale can be much smaller than what is expected from Coulomb interaction alone (e.g. \citealt{Vink15}). 
Following from the energy density of the relativistic particles, a non-thermal SZ contribution from the power-law electrons should also be very small \citep{EK00} .}

An important point is that the SZ surface brightness is redshift independent, whereas in contrast, the \syn flux rapidly decreases with redshift, both for the redshift dimming and the strong $K$-correction to the observed flux due to the steep \syn spectrum of the relics. This implies a rapid increase in the SZ flux contamination for clusters at redshift $z\sim 0.5$ or above, compared to mostly low-$z$ clusters in which radio relics are observed today (Section \ref{sec:zdep}). Not only for radio relics, this redshift dependence of the SZ flux contamination will also play a critical role for GHz-frequency observation of radio halos (Section \ref{sec:szhalo} below). This is important for the large number of cluster diffuse sources (relics and halos) that are expected to become available from upcoming radio surveys with ASKAP, MeerKAT and SKA.

%%%%%%%%%%%%%%%%%%%%%%%%%%%%%%%%%%%%%%%%%%%%%%%%%%%%%%%%%%%%%%%%%%%%%%%%%%%%%%%%%%%%%%%

\subsection{Relic \rev{synchrotron} emission at the shock front}
\label{sec:radio}

In the first-order Fermi process, the momentum distribution of the accelerated particles follow a power-law, $f(p) \propto p^{-\delta}$, and the slope of the particle spectrum can be related directly to the shock compression ratio, $r$,
\begin{equation}
\delta = (r+2)/(r-1)
\label{eq:delta}
\end{equation}
This particle distribution produces synchrotron radiation also with a power-law: $S(\nu) \propto \nu^{-\alpha_{\mathrm{inj}}}$, where $\alpha_{\mathrm{inj}} = (\delta-1)/2$ is generally termed as the injection spectrum for the relics. But this is only observable at the very leading edge of the relic, where energy losses are negligible. In practice, the synchrotron spectral index is generally measured over the entire relic area, where particle ageing as well as continuous injection of new particles produce a power-law which is steeper: $\alpha_{\mathrm{tot}} = \alpha_{\mathrm{inj}} + 0.5$ (\citealt{Kard62}, \citealt{Enss98}). We shall use this integrated spectral index for the rest of our work, simply assuming that the brightness profile of the relic retains its shape at all frequencies. Putting in $\gamma=5/3$ in the formula for compression ratio, we get the Mach dependence of the \syn spectral index
\begin{equation}
\alpha_{\mathrm{tot}} \left(= \delta/2\right) = (\mach^2+1) / (\mach^2-1)
\label{eq:alpha}
\end{equation}

It can be seen that the spectral index is a weakly varying function of the Mach number for $\mach \gtrsim 2$, weak shocks producing a steeper radio spectrum and hence less flux at cm-wavelengths. 
However, a more pronounced $\mach$ dependence can be expected for the amplitude of the radio power at any given frequency, since the kinetic energy dissipated by the shock scales as the third power of shock velocity downstream, hence as $\mach^3$. This is more pronounced than the roughly $\mach^2$ increase of the SZ signal due to pressure jump (Eq. \ref{eq:Pratio}). We shall come back to the $\mach$ dependence of the SZ contamination in Sec. \ref{sec:machdep} to illustrate the concept qualitatively. More  accurate computation of the \syn power dependence on the Mach number will require some degree of modeling, because the particle acceleration efficiency and magnetic field amplification by weak cosmological shocks are mostly unknown (e.g. \citealt{HB07}, \citealt{Pinz13}). However, for our work we can avoid these complications, by using either the measured flux values of known radio relics, or using some empirical scaling relation between the cluster mass and relic power.

The last piece of information that we need to predict the flux contamination in a realistic manner is the shape of the \syn brightness profile. Here, lacking more detailed simulations of CRe production and magnetic fields, we employ an empirical model that matches actual measurements. We find that a lognormal emissivity profile behind the shock front produces acceptable fits to the published relic flux profiles, \rev{that can also be modified to fit for the unknown viewing angles of the relics,} as described below.

%%%%%%%%%%%%%%%%%%%%%%%%%%%%%%%%%%%%%%%%%%%%%%%%%%%%%%%%%%%%%%%%%%%%%%%%%%%%%%%%%%%%%%%

\subsection{A~``lognormal model'' for the \rev{synchrotron} profile}
\label{sec:lognormal}

To make realistic predictions for the flux contamination, we need a \rev{synchrotron flux distribution} model that can match the observed data at low-frequencies as closely as possible, and which is scalable to the basic relic properties like its length (largest linear scale, or LLS, of the relic) and the shock Mach number. 
%Due to particle ageing, the synchrotron emissivity power-law steepens behind the shock front, 
%follows a power-law: $S(\nu) \propto \nu^{-\delta/2}$, where $\delta$ is the particle energy spectrum. Thus 
A crude emissivity model will consist of a sharp rise \rev{of the synchrotron emissivity} at the shock front, followed by a power-law decrease downstream \rev{as the particles lose energy}. The reasons for choosing a log-normal distribution instead are {\it (i)} it makes a more physical change at the shock front that is non-instantaneous, and {\it (ii)} it offers a simple formulation of the relic width which we can connect to the particle energy spectrum. This is still a distribution whose tail is flatter than exponential, approximating to a power-law beyond the typical width of the shock. The functional form for the flux radial distribution is then 
\begin{equation}
S_{\nu}^{\mathrm{sync.}}(x_d) ~ = ~S_0 ~\left(\dfrac{e^{\displaystyle{-(\log x_d - \mu)^2 / 2\sigma^2}}}
{\sqrt{2\pi} ~ x_d ~ \sigma}\right) ~~~ \mathrm{(mJy/arcmin}^2)
\label{eq:lognorm}
\end{equation}

The normalization $S_0$  in Eq. \ref{eq:lognorm} above is fixed to match real data \rev{(e.g. Section \ref{sec:indiv})}, by integrating the profile above a certain significance level \rev{(typically $\sim 20\sigma$)}, and multiplying by the relic length and the mean flux to match total flux values. The two parameters $[\mu, \sigma]$ for the shape fitting denote the location and scale of the distribution, and $x_d$ is the downstream length in kpc. There can be two viable methods to obtain the fit: we can either fit the {\it projected} emission profile directly using a lognormal model, or assume a {\it 3D} emissivity profile that follow lognormal distribution and fit the data after projection. We find that at least for edge-on relics both methods can offer reasonable fit to the data, but we prefer the latter method because it  allows for a natural way to make a rotation and view the relic from arbitrary angles. 

\begin{figure}[t]
%\centering
\includegraphics[width=\columnwidth]{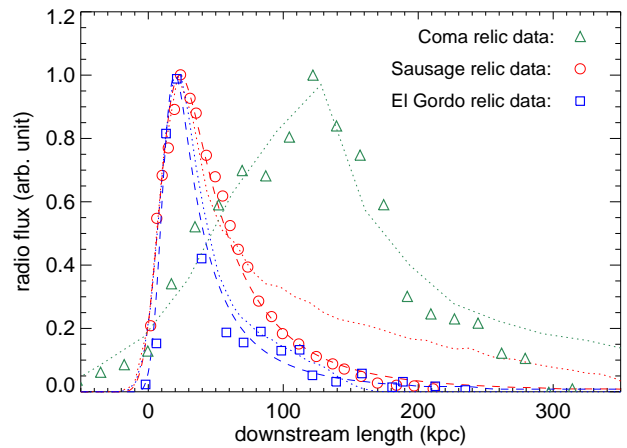}
\caption{Fit to the observed flux profiles for three radio relics with our lognormal model. The three different colored symbols represent the flux profile data points in units of physical length, and the dashed and dotted lines with corresponding colors show model fits.  
The data point are from van Weeren et al. (2010) for the Sausage relic, Lindner et al. (2014) for the El Gordo NW relic, and Trasatti et al. (in prep.) for the Coma relic, all normalized to the same amplitude. For the roughly edge-on relics (Sausage and El Gordo), we show results for two different methods of obtaining a fit: one by directly fitting the projected brightness profile in 1D (dashed lines), and the other by using a 3D emissivity model and fitting through projection (dotted lines). The second method produces a long emission `tail' for the Sausage relic. For the Coma data only a 3D result is shown, which provides acceptable fit after accounting for a $\sim 30^{\circ}$ viewing angle (see text). The models are convolved with the respective instrument resolutions as noted in the publications. The shock location is at zero, cluster center is to the right.} 
\label{fig:lognormfit}
\end{figure}

The results for fitting \rev{the lognormal model of Eq. \ref{eq:lognorm}} to three sets of relic data are shown in Fig. \ref{fig:lognormfit}. The flux profile for the \sausage relic is taken from \citet{vanWe10}, for the NW relic in El Gordo from \citet{Lind14}, and the Coma relic data from Trasatti et al. (in prep., see \citealt{Er15}). We normalize each data peak to unity, and have used the published instrument resolutions to \rev{beam-convolve our profiles.}  
When fitting the edge-on relics directly to their on-sky brightness profiles, we require parameter values roughly as $[\mu, \sigma] = [3.2, 0.8\alpha_{\mathrm{tot}}]$, whereas fitting the volume emissivity profile require much narrower distribution, roughly $[\mu, \sigma] = [0.1, 0.3\alpha_{\mathrm{tot}}]$ \rev{(these values are for the three cases only, and should not be considered as general relic properties).}  
The projection provides the requisite broadening in the 3D case, \rev{and hence the underlying lognormal profile is narrower.} For the \sausage relic we note an excess emission ``tail" in the downstream direction, which is inevitable due to our assumed circular geometry if the relic diameter (i.e. its LLS) is large (1.7 Mpc for \textit{Sausage}). Therefore, for  specific cases like the \sausage relic we fit the projected brightness for better accuracy, although our results will not change significantly when using a 3D emissivity model. For the Coma data an edge-on relic  can not provide a good fit, so we use the 3D emissivity model with a viewing angle of roughly $\sim 30^{\circ}$ (Fig. \ref{fig:lognormfit}). Similarly, large viewing angle is needed for fitting the relic data in A2256, as we discuss in Sec. \ref{sec:a2256}. We note that there is significant degeneracy between the lognormal model parameters with the viewing angle, so our fit results should not be considered as unique solutions to these systems, but rather an approximation to match the observed radio data realistically.

\begin{figure}[t]
%\centering
\includegraphics[width=0.9\columnwidth, height=4.6cm]{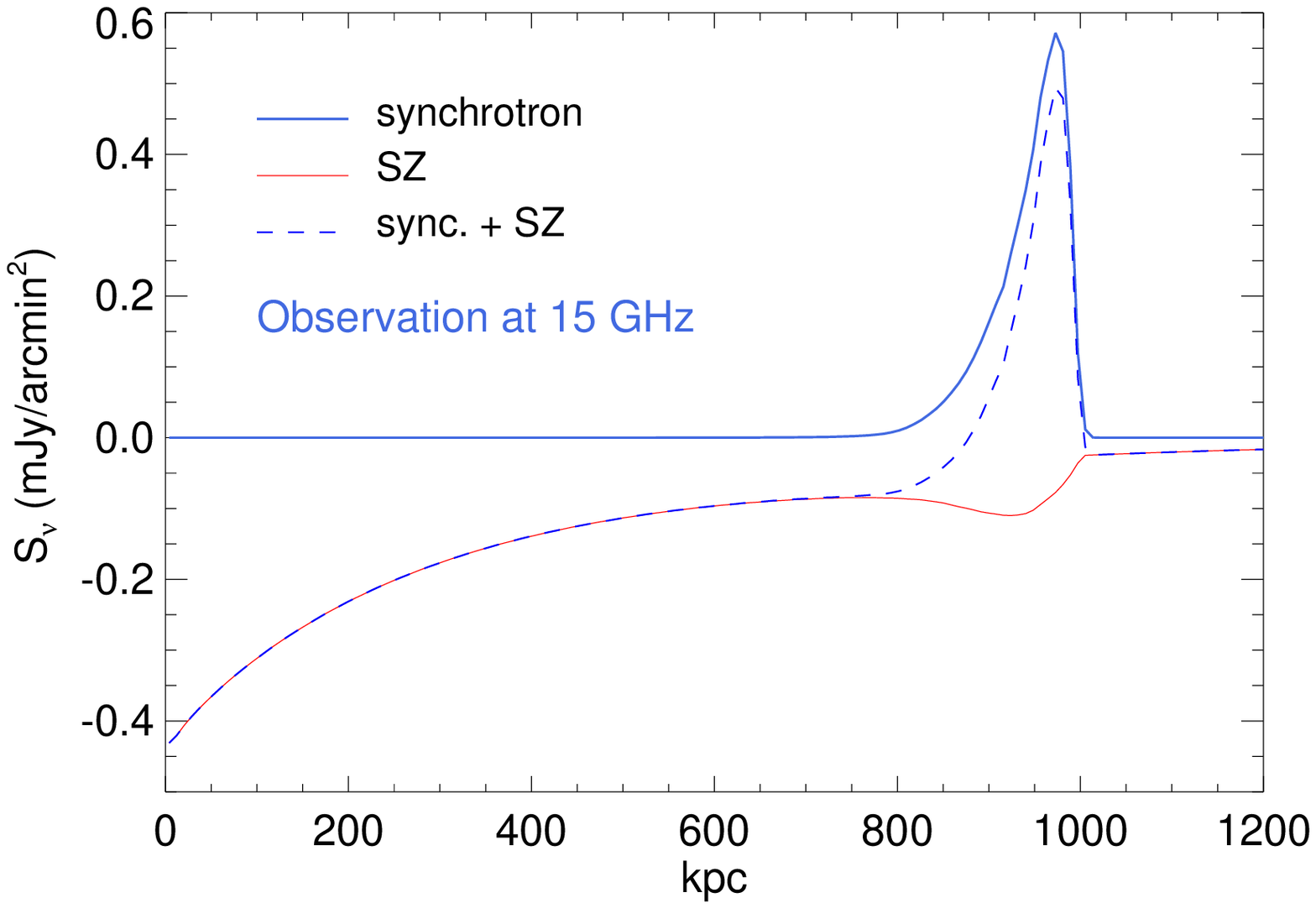}
\includegraphics[width=0.9\columnwidth, height=4.5cm]{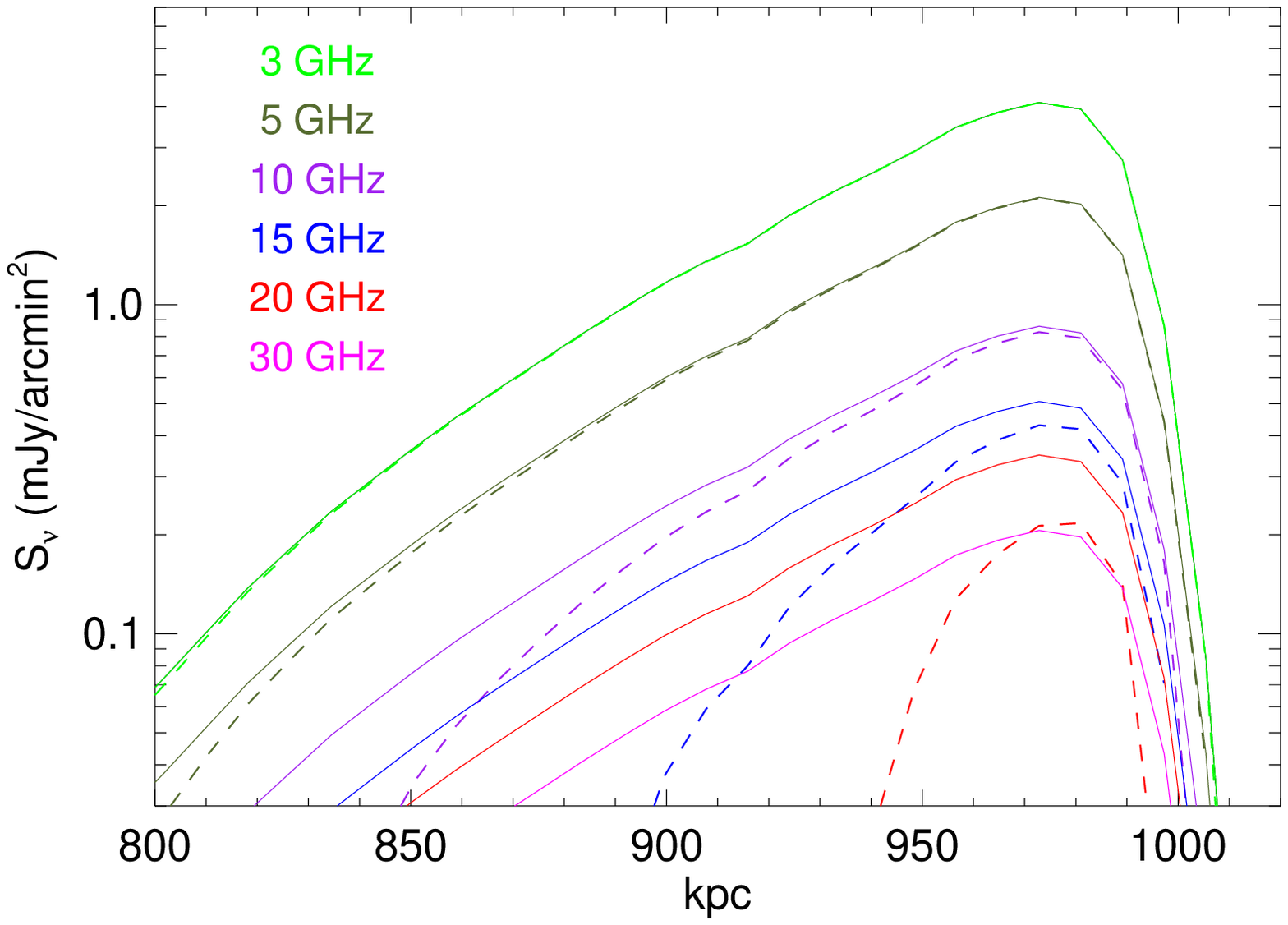}
\caption{
Illustration of the SZ and \syn flux profiles across the location of a relic shock at 1 Mpc. The construction of the SZ and \syn fluxes are described in Sections \ref{sec:sz} and \ref{sec:lognormal}, respectively, and the example cluster is similar to the one with \sausage relic. In {\it top panel} we show the case at a single frequency (15 GHz), for the full projected length scale of the cluster.  The blue dashed line mark the modified radio flux. 
In the {\it bottom panel} the true and observed \syn fluxes are shown in solid and dashed lines in log-scale. At 30 GHz the signal is negative throughout due to the SZ decrement, although interferometric observation will be able to recover some positive signal (see Sec. \ref{sec:interferometry}). 
For relics with significant inclination angles the \syn profile is modified while keeping the SZ unchanged (Sec. \ref{sec:angle}).}
\label{fig:profiles}
\end{figure}

A comparison for the relative amplitudes of the SZ and \syn fluxes is shown in Fig. \ref{fig:profiles} top panel, for a fiducial shock model in a cluster of mass $6\times 10^{14}$ M$_{\odot}$ at $z=0.2$, shock Mach number $\mach = 3$ and shock radius 1 Mpc.  The \syn profile is obtained through projection of a 3D edge-on relic model with lognormal emissivity profile, and matched to a total \syn power $P_{1.4} = 4\times 10^{24}$ W/Hz. 
In the bottom panel of Fig. \ref{fig:profiles} positive sections of the modified \rev{synchrotron}$+$SZ flux profiles are shown in dashed-lines, which we compare with the true \syn signal when computing contamination values. 
One simplifying assumption in our modeling is that the radio flux profile retains its shape at all frequencies, i.e. we ignore the spectral index variation across the relics and only use a global mean value, as this assumption does not change our results in any major qualitative way. 

\begin{figure}[h]
%\centering
\includegraphics[width=\columnwidth]{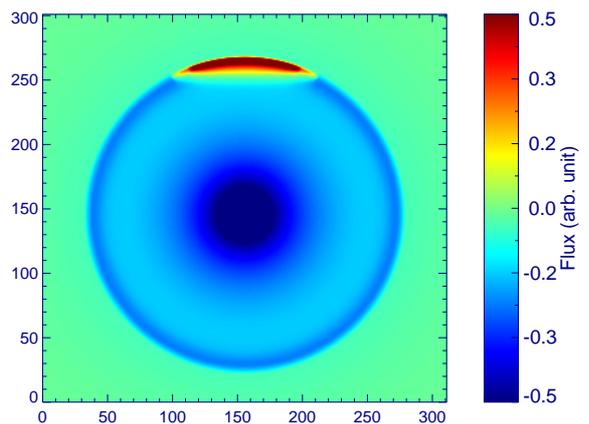}
\caption{Example of a 2D image used for interferometric simulation. The \syn and SZ emissivity models are placed on a spherical shock and projected, on top of a cluster-wide SZ signal that follows the GNFW pressure model. In this case the relic is viewed edge-on ($0^{\circ}$ inclination angle). The full-spherical shock model for SZ is obvious, although for our purposes of computing flux contamination it is sufficient for the SZ signal boost to be localized in the same angular segment as the radio relic emission.}
\label{fig:2Dimg}
\end{figure}

Fig. \ref{fig:2Dimg} presents an example of how the \syn and SZ signals appear together after projection. The lognormal emissivity model follows the same 3D spherical shock geometry as discussed in Sec. \ref{sec:sz} for computing the SZ signal.   
The crucial distinction from the SZ shock model is that \syn emission is localized within a circular patch whose diameter is determined by the measured relic linear scale (LLS). It is an observational fact that radio relics are localized emissions, even rare objects like the \sausage relic do not trace a complete spherical shock front. One possibility is that this is due to a most efficient electron acceleration in patches of magnetic fields that are perpendicular to the shock normal (e.g. \citealt{Guo14a}, \citeyear{Guo14b}).  In fact, for relics that are viewed edge-on it is sufficient to have the SZ shock to be localized in the same region as marked by the \syn signal, since in projection the SZ signal boost gets its contribution mostly from a small region near the edge of the shock front. For relics viewed at large inclination angles the SZ signal modeling is generally more uncertain as the actual geometry of the shock front will have more relevance.

%%%%%%%%%%%%%%%%%%%%%%%%%%%%%%%%%%%%%%%%%%%%%%%%%%%%%%%%%%%%%%%%%%%%%%%%%%%%%%%%%%%%%%%
\subsection{Side topic: SZ contamination for radio halos}
\label{sec:szhalo}

Unlike radio relics, the issue of SZ flux contamination has already been considered in the context of radio halos (\citealt{Enss02}, \citealt{Pfrom04}, \citealt{Bru13}). 
They have similar steep \syn spectrum as in relics, so at GHz-frequencies their observed flux will likewise be affected by the SZ effect. Here the standard picture of ``negative bowl'' of SZ effect applies: radio halos sit near the cluster centers where the SZ signal is strongest, and both of these signals can have similar radial profiles \citep{PComa}. The highest frequency measurement of a radio halo is at $\sim 5$ GHz, for the nearby Coma cluster at $z=0.02$ \citep{Thi03}. \citet{Bru13} have argued that at 5 GHz the effect of SZ decrement is negligible, accounting for less than 10\% of the total \syn flux at that frequency and hence can not be responsible for the observed steepening of the radio spectrum. Using the GNFW model (eq. \ref{eq:gnfw}) fit for Coma \citep{Er15}, we can confirm that the SZ contamination is indeed $\lesssim 10\%$ at 5 GHz, so no current radio halo data has been seriously affected by this issue.

However, due to the redshift independence of SZ brightness, the flux contamination for radio halos will increase rapidly with redshift. We consider the famous El Gordo cluster (ACT-CL J0102$-$4915) as an example, which at $z=0.87$ is currently the highest redshift radio halo known. Its highest frequency radio measurement is at 2.1 GHz \rev{\citep{Lind14}}, and using the GNFW pressure model appropriate for the mass and redshift for this cluster, we find a flux contamination value of only 4\% within the radio halo radius ($r_{\mathrm{H}}$ is similar to $r_{500}$, roughly 1.2 Mpc). If El Gordo radio halo is observed at 5 GHz as in Coma, the contamination will increase to 55\%, i.e. observed \syn flux would be roughly half of the true flux (SZ flux roughly $-0.5$ mJy compared to extrapolated synchrotron flux $0.9$ mJy at 5 GHz, \rev{using a  spectral index $\alpha=1.2$}).  
We obtain similar high contamination values at 5 GHz for other high-$z$ radio halos/mini-halos. 
For the  Phoenix cluster at $z=0.6$ \citep{vanWe14a}, contamination is roughly 13\% within the small mini-halo radius of $\sim 500$ kpc. For the giant radio halo in PLCK G147.3$-$16.6  at $z = 0.65$ ($r_{\mathrm{H}} \sim 1$ Mpc; \citealt{vanWe14b}), the flux contamination at 5 GHz will be around 37\%. 
We do not elaborate further on this issue of radio halo flux contamination, but rather return to radio relics and their flux measurements.

%%%%%%%%%%%%%%%%%%%%%%%%%%%%%%%%%%%%%%%%%%%%%%%%%%%%%%%%%%%%%%%%%%%%%%%%%%%%%%%%%%%%%%%
\subsection{Interferometric and single-dish observations}
\label{sec:interferometry}

\rev{Radio data are obtained from either interferometric or single-dish observation.} The advantage of interferometers is their high angular resolution, which helps to remove point source contamination from the relic images. Therefore, also for single-dish imaging it is desirable to have complimentary interferometric data \rev{for separating the compact emission}. Interferometers, on the other hand, will  miss some large-angle flux, on scales larger than $\sim \lambda/D_{\mathrm{min}}$, where $D_{\mathrm{min}}$ is the smallest baseline length. 
For cm-wavelength observation \rev{(1--30 GHz)} this is particularly a problem, as there are not many interferometers with short enough baselines that can image scales of several arcminutes, as is often exhibited by radio relics. 
Indeed, the two interferometers that have been used for cm-wavelength relic imaging (AMI and CARMA) were specifically designed to image the cluster-wide SZ signal at these wavelengths. 
Loss of signal due to a lack of sufficiently short spacings can be a problem for all known relics at cm-wavelengths, save for compact high-$z$ sources like the El Gordo relic \rev{(whose length is $\lesssim 1$ arcmin)}.  
Single-dish measurements have mostly been done by the Effelsberg 100 m radio telescope, using its $3-10$ GHz bands (\citealt{Tra15}, \citealt{Str15}).

The basic difference between single-dish and interferometric imaging is that the former measures total power (depending upon the scan length), whereas the latter is sensitive to the signal autocorrelation in the plane of the sky, i.e. measures the change in brightness. It might be assumed that since the SZ signal is generally large-scale compared to the localized \rev{synchrotron} emission, the effect of interferometric imaging would be to remove the large-scale SZ flux while keeping the \rev{synchrotron} part intact, thereby recovering the true \rev{synchrotron} signal (assuming the relic to be sufficiently compact). While this would be the case if the radio relic was situated somewhere in the negative ``bowl'' of the cluster-wide SZ signal, the shock also creates a sharp SZ flux variation on the same scale as radio relics (top panel of Fig. \ref{fig:profiles}), 
thus single-frequency interferometric imaging will not be able to disentangle the \rev{synchrotron} signal from the SZ component.  
How much of the radio relic flux is actually recovered depends upon the details of the image deconvolution process, the {\it uv}-coverage in relation to the scale of the emission, and the signal-to-noise ratio of the measurement.

\begin{figure}[t]
\centering
\includegraphics[width=\columnwidth]{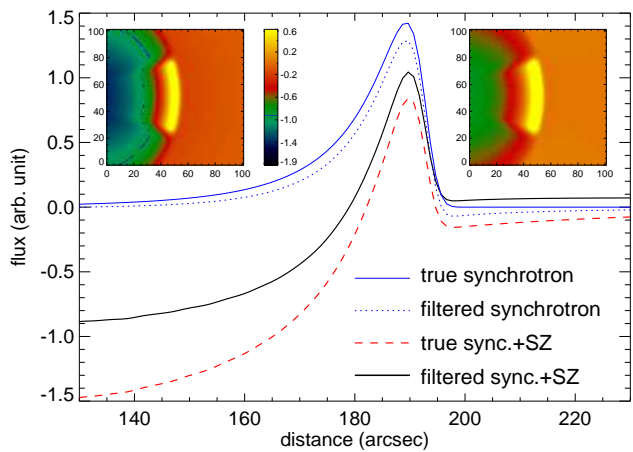}
\caption{High-pass filtering of a 2D radio relic model in the presence of SZ decrement, done as a simplified exercise for interferometric observation. The inset panels show the input image before {\it (left)} and after {\it (right)} filtering, using the same color scale. The multiple curves in the main panel plot represent the profiles for: \rev{synchrotron}-only model (blue solid line), filtered \syn model (blue dotted line), \rev{synchrotron}$+$SZ model (red dashed line), and filtered \rev{synchrotron}$+$SZ model (black solid line). We see that the high-pass filter is effectively removing the large-scale SZ signal from the image, while keeping the \rev{measured peak flux close to the input (unfiltered) value of the \rev{synchrotron}$+$SZ model, and not the `\syn only' model.}}
%keeps the peak of the modified radio signal close to its input value, rather than the value corresponding to the radio-only model.
\label{fig:hipass}
\end{figure}

\subsubsection{High-pass filtering of the total signal}

Let us first consider a simple exercise of high-pass filtering of a noiseless radio relic image, in the presence of negative SZ signal, to get an intuitive understanding of imaging with interferometers. The input image is a simplified radio \syn relic plus SZ model, shown in the left inset of Fig. \ref{fig:hipass}. Onto this we apply a high-pass filter (Butterworth filter of first order), whose cutoff frequency is designed to keep the original relic signal mostly unchanged. The image after filtering is shown in the right inset with the same color-scale. The radial profiles  before and after the high-pass filtering is shown in the main panel of  Fig. \ref{fig:hipass}. Here the blue solid line is the \syn signal without any SZ effect, and the red-dashed line is the \rev{synchrotron}$+$SZ signal that an ideal total-power imager will find. The filtered profile is shown by the solid-black line, that will be the result of an interferometric observation. A filtered \rev{synchrotron}-only profile is shown in blue-dotted line, which is lower than the input \syn model. 
 
Three things are to be noted from this exercise of high-pass filtering. First, the large-scale SZ signal is mostly removed from the filtered image, and the peak of the measured signal (black solid line in Fig. \ref{fig:hipass}) corresponds closely to the modified \syn signal (red dashed line), meaning \rev{SZ-shock induced flux modifications are there also after filtering.} 
Second, the filtering would also affects the original \syn signal to some extent (blue dotted line), meaning there would be some flux loss also without SZ contamination.  
Third, and more subtly, the filtering process will ``compensate'' for some of the flux loss by removing the large-scale negative SZ component, as expected from a high-pass filter (more positive flux under the black solid line than the red dashed line). As we show in the next subsection, the situation is somewhat altered when we consider a more realistic simulation of an interferometric observation.

\subsubsection{Simulated interferometric observation}

In terms of radio interferometry, a high-pass filtering corresponds, \rev{to a first approximation}, to an  inverted (or ``dirty'') image in presence of infinite signal-to-noise and full coverage of the unfiltered part of the $uv$-plane. 
\last{In the case of real observations, one typically uses the CLEAN algorithm to do the imaging, ``filling in'' the missing information in the $uv$-space by interpolating from the existing data, thus producing an image with minimal artifacts (see \citealt{RIbook99} for details).} 
In a radio interferometric observation, flux can also be recovered on scales much larger than what is expected from a simple consideration of the shortest baseline. For example, if one dimension of the source is much larger than the other (as in most radio relics), a mosaic of several \rev{pointed observations} can recover the full flux of the source, given that the smaller dimension is within range of the interferometer array. However, such signal reconstruction is satisfactory only in the limit of very high signal-to-noise, and an almost complete $uv$-coverage (full rotation synthesis). 

\rev{The reconstruction of the modified relic signal is more complicated, as there will be both positive and negative signal components. For moderate signal-to-noise cases (cm-wavelength relic observations typically have S/N $\sim 5$ for the peak flux), one would generally apply the CLEAN algorithm by selecting regions (``clean boxes'') around sources.  In case of radio relics it can be natural to select regions which only show a positive residual flux in a noisy field. Thus the outcome of the image deconvolution  will vary depending on the CLEAN regions applied, as well as the relic shape and size and the signal-to-noise of the measurement.} 
%on the positive residual signal, and outcomes will vary on the implementation of this technique as well as the relic shape and size and the signal-to-noise of the measurement.

To make realistic simulations for interferometric imaging, we employ the CASA\footnote{http://casa.nrao.edu/} software package, in particular its \textsc{simobserve} task, to create a measurement set from the model image (similar to Fig. \ref{fig:2Dimg}),  
%which is then CLEANed interactively. 
which is then deconvolved using CLEAN with the Briggs weighting (robust=1) scheme. The clean mask is defined from the signal-to-noise ratio of the residual image, cleaning only in regions where this ratio is greater than five. 
These parameters are chosen to minimize the flux loss from the diffuse relic signal, and cleaning is performed \rev{interactively}. 
We simulate a VLA X-band (10 GHz) observation in D-configuration, and considering the limited spacial fidelity of VLA at these frequencies \rev{(largest recoverable scale $\sim 2$ arcmin)}, we make a compact relic model similar to the NW relic in the El Gordo cluster, such that the angular extension of the relic is roughly 1.5 arcmin and one would not need  mosaicing. \rev{The thermal noise level from the simulator is chosen to have a peak S/N $\sim 10$ in the image}. We simulate both a `synchrotron-only' model as well as a `synchrotron plus SZ' model, with the same noise realization, to compare the fluxes after CLEANing. 
Primary beam correction is taken care of by the \textsc{simobserve} task.

\begin{figure}[ht]
%\centering
%\hspace*{-5mm}
\includegraphics[width=0.65\columnwidth]{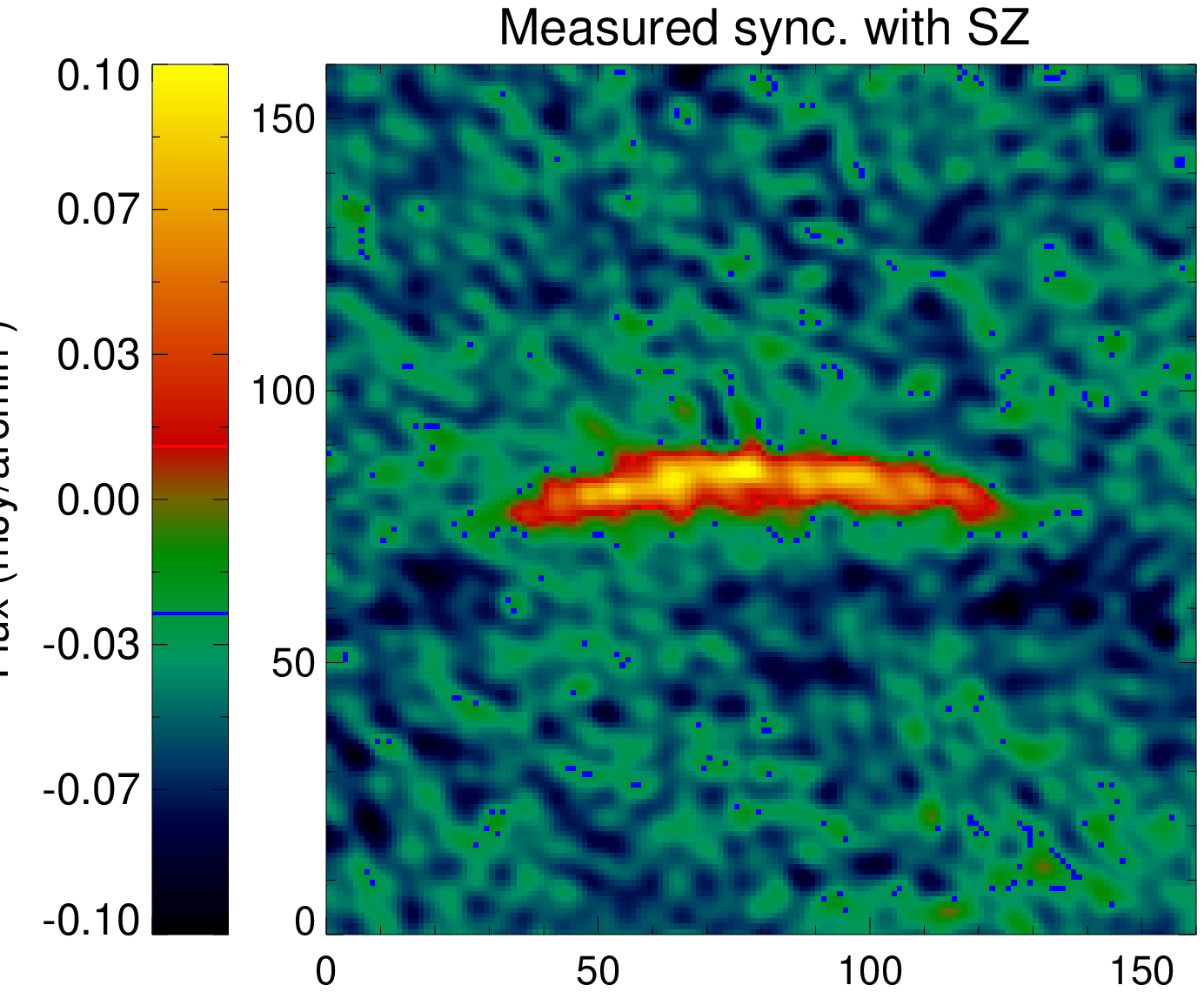}%
\hspace*{-18mm}
\includegraphics[width=0.65\columnwidth]{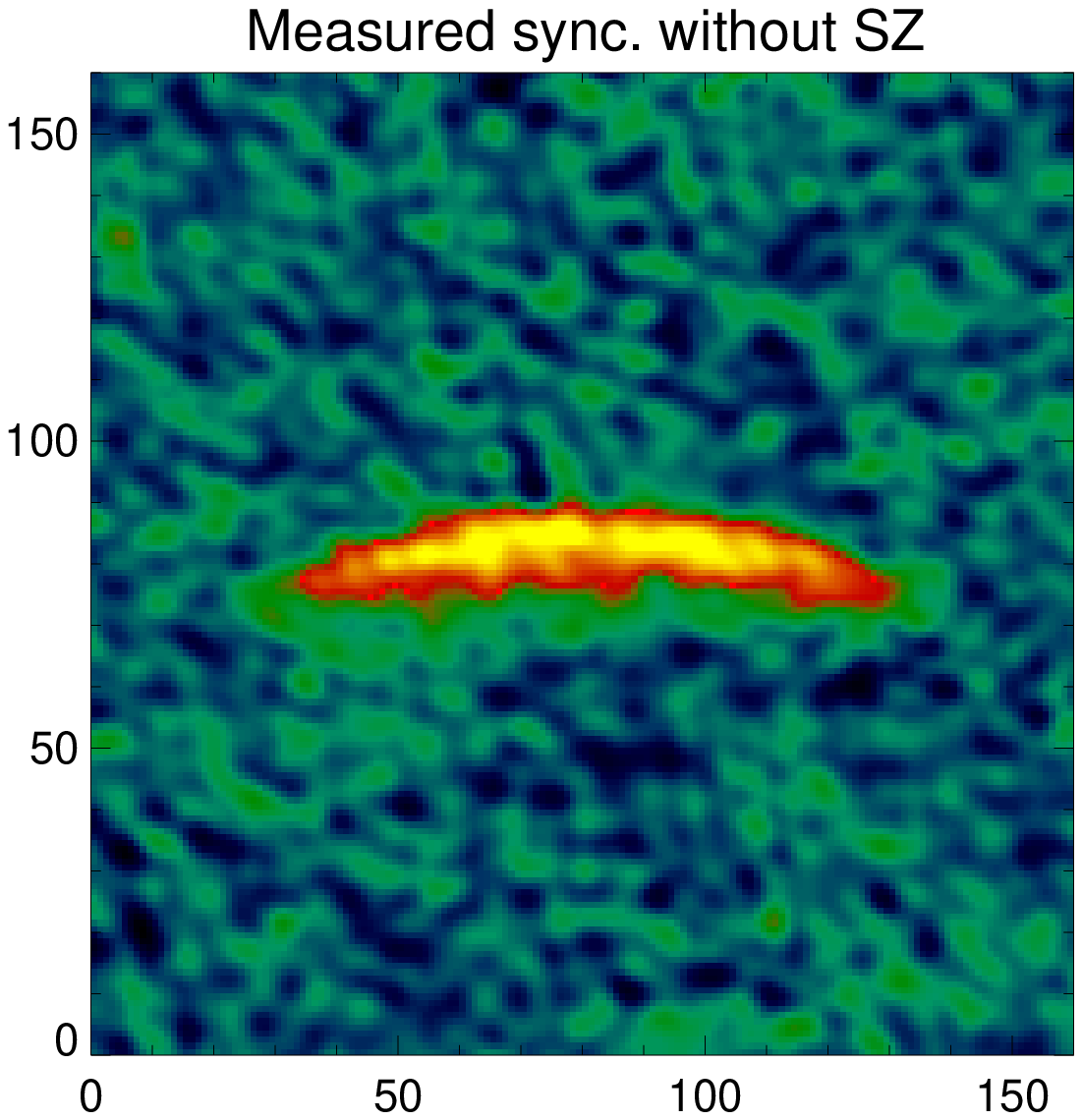}
\hspace*{-3mm}
\includegraphics[width=0.98\columnwidth, height=5.7cm]{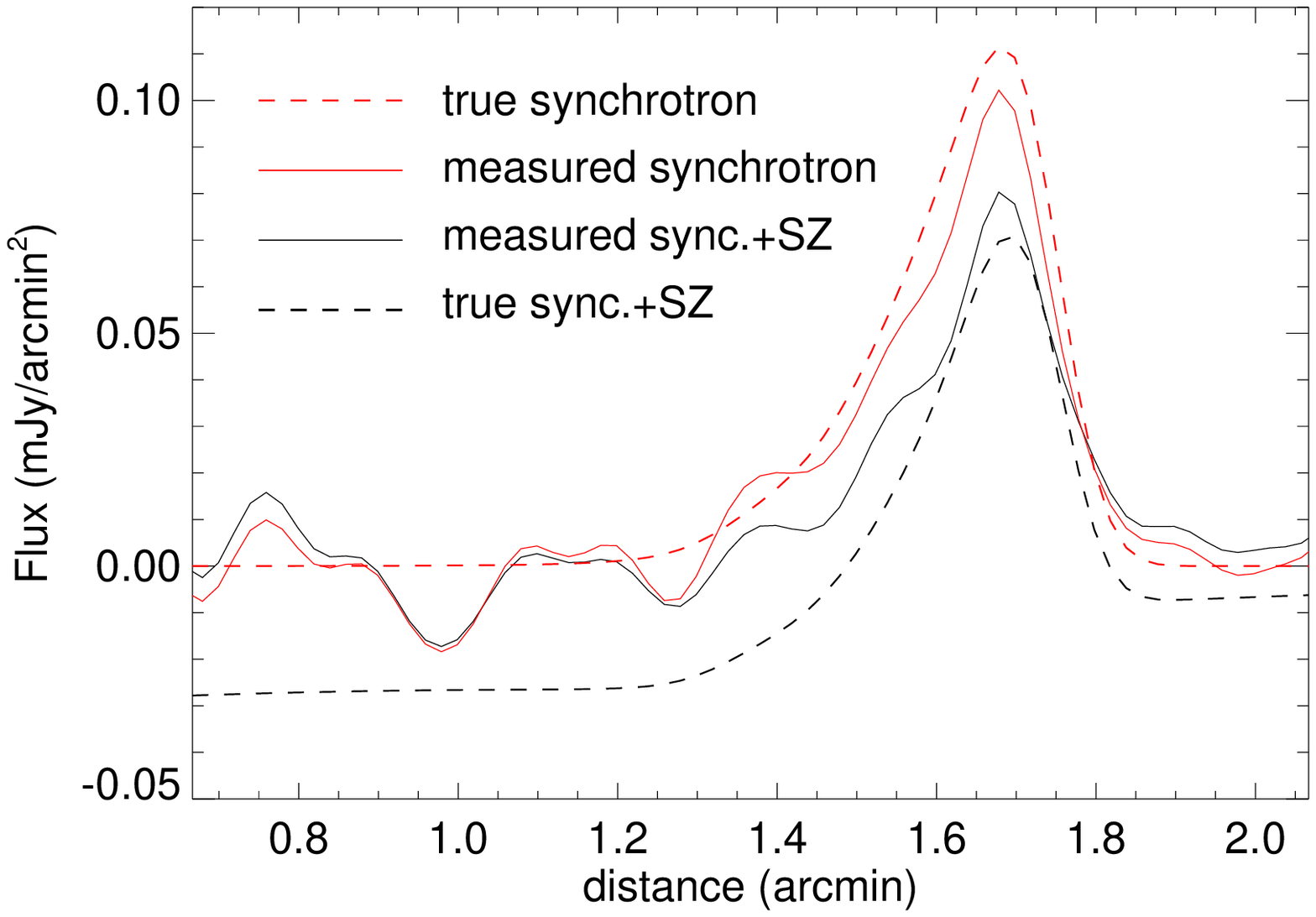}
\caption{Simulated interferometric observation at 10 GHz for an El Gordo-type compact relic with the VLA X-band. Simulations are done by the CASA software and CLEANed interactively. {\it Top panel left:} Actual measurement of the radio relic in presence of SZ flux contamination. {\it Top panel right:} Hypothetical measurement of only the \syn emission, without any SZ. \rev{The image axes are in arcseconds, both cases use the same noise realization, and are shown in the same color scales.} {\it Bottom panel:} The flux profiles before and after processing through the simulation set-up. The red-dashed line is the input \rev{synchrotron}-only model, and the black-dashed line is the \rev{synchrotron}$+$SZ input model. Red-solid line shows the effect of interferometer on a \rev{synchrotron}-only flux measurement, which also registers some flux loss due to lack of short baselines. The black solid line is the actual observed flux with interferometers when SZ is also taken into account.}
\label{fig:intsim}
\end{figure}

The simulated observations are shown in Fig. \ref{fig:intsim}. 
%We show both the original radio model and the radio$+$SZ model results for comparison, which were simulated with the same noise model. 
The top panels show the CLEANed images of radio \syn signal in presence of SZ flux decrement (left), and only the radio \syn signal (right). The flux loss in the presence of SZ is evident, both from a reduced peak value and a shrunken width. The curves in the lower panel show the observed flux profiles in comparison to the input models. The black-dashed line is the input model of \syn plus SZ, that will be observed by an ideal total-power instrument. The actual interferometric observation is the black solid line, missing the SZ negative parts completely and yielding more positive flux than the model. This is to be compared with the input \rev{synchrotron}-only model (red-dashed line), which also suffers from flux loss due to a lack of sufficient short baselines. To illustrate by numbers: the total flux in our input \syn model at 10 GHz is 0.22 mJy, and the total positive flux in presence of SZ decrement is 0.11 mJy. So the flux contamination is 50\% for an ideal observation 
%(see Table \ref{onetable} for contamination definition). 
%\rev{(contamination is defined as the ratio between lost flux and the true flux).} 
\rev{We define the contamination fraction as the ratio between the flux lost due to the SZ effect and the true (uncontaminated) synchrotron flux. Thus the contamination (in percentage) is $C = 100\times (\mathrm{true~ flux} - \mathrm{measured~ flux})/\mathrm{true~ flux}$, which makes it roughly equivalent to the flux ratio: $C \approx 100\times (\mathrm{SZ ~flux/synchrotron ~flux})$. }
In the case of simulated  interferometric imaging with CLEANing, the recovered \rev{synchrotron}-only flux is 0.18 mJy, and the \syn flux in presence of SZ contamination is 0.13 mJy. So the interferometric observation produces a flux contamination roughly 40\%, somewhat less than our idealized value. This flux change is due to both the presence of SZ signal (majority contribution) as well as loss of signal due to missing short spacings for the faint and extended source.

To summarize, interferometric imaging of radio relics at cm-wavelengths will register flux loss \rev{similar to a single-dish measurement}, due to the presence of a relic-scale negative SZ feature from the shock, and also -- by a smaller amount -- due to the insufficient $uv$-coverage when imaging a large, diffuse relic emission. \rev{From the particular case analyzed with our mock interferometric observation, we found that the  combined effect of these two modifications can be roughly equal to the diminished flux measured by an `ideal' total-power instrument, that registers both the positive and negative fluxes accurately.} 
%The combined effect of these two changes is close to the result of an `ideal' total-power instrument that registers both the positive and negative fluxes accurately. 
We can assume that flux losses due to inadequate short baselines an almost inevitable feature for high-frequency radio imaging, unless compact, high-$z$ relics are targeted or special care is taken to combine interferometric results with single-dish data (`zero-spacing' addition). The actual amount of flux that is recovered by any specific interferometer is a complicated function of signal-to-noise, the source size and shape, amplitude of the ambient SZ signal, etc, and is difficult to predict. It will also not scale as a fixed fraction of the ideal contamination number, due to the changing shape of the flux profile with frequency. Therefore, in our presentation of the contamination results in this paper, we quote contamination values simply by comparing the true positive \rev{synchrotron}-only fluxes with the measured positive radio fluxes after adding the SZ decrement.

%%%%%%%%%%%%%%%%%%%%%%%%%%%%%%%%%%%%%%%%%%%%%%%%%%%%%%%%%%%%%%%%%%%%%%%%%%%%%%%%%%%%%%%
\section{Results}
\label{sec:allresults}
%%%%%%%%%%%%%%%%%%%%%%%%%%%%%%%%%%%%%%%%%%%%%%%%%%%%%%%%%%%%%%%%%%%%%%%%%%%%%%%%%%%%%%%

In \rev{the first Section \ref{sec:trend}}  we present some mean trends for the flux contamination: namely how it varies with redshift, cluster mass, shock Mach number, and also the relic viewing angle. \rev{In the following Section \ref{sec:indiv}} we provide results for some well known relics in individual clusters. \rev{Results from the latter section are summarized in Table \ref{onetable}.}
%When we quote a flux contamination number, it is computed from the ratio of the measured positive flux against true positive flux (that one would have observed in the absence of the SZ effect). Thus the contamination (in percentage) is $C = 100\times (\mathrm{true~ flux} - \mathrm{measured~ flux})/\mathrm{true~ flux}$, which makes it roughly equivalent to the flux ratio: $C \approx 100\times (\mathrm{SZ ~flux/radio ~flux})$. 
%(see Table \ref{onetable} and Sec. \ref{sec:theory}). 

%\subsection{\textit{\textbf{Part I. Mean trends for the SZ contamination}}}
\subsection{\textit{\textbf{Modeling results for the mean contamination}}}
\label{sec:trend}

First we use the GNFW cluster model (Eq.\ref{eq:gnfw}) and an empirical scaling relation between a cluster's mass and its radio relic power, to predict the general trend of the SZ flux contamination with cluster mass and redshift. We also describe, using simplified models, how the SZ-to-\syn flux ratio is expected to vary with shock Mach number, and use the rotation of the relic model described in Section \ref{sec:lognormal} to qualitatively outline the impact of relic viewing angles.

\subsubsection{SZ contamination as function of redshift}
\label{sec:zdep}

As the SZ surface brightness is practically redshift independent, the flux contamination will increase rapidly with redshift. What is somewhat unexpected is that for higher mass systems we get less contamination. This is because of the strong mass dependence of the radio relic luminosities that is found empirically ($P_{1.4} \propto M_{500}^{2.8}$; \citealt{deGas14}), which is steeper than the mass dependence of the cluster SZ signal ($Y_{\mathrm{SZ}} \propto M_{500}^{5/3}$). Assuming the shock Mach number and magnetic field intensity to be independent of cluster mass, this will mean that for lower mass systems the \syn power will fall off more quickly than the SZ signal, causing more significant contamination. We do not attempt to explain this 
 strong mass dependence of radio relic power, see \citet{deGas14} for a discussion on this issue. However, we note that similar strong mass dependence is also found for radio halos \rev{(e.g. \citealt{Som14})}, and there too the SZ flux contamination should decrease with increasing mass.

We fix the normalization of the empirical scaling result from \citet{deGas14} as following: 
\begin{equation}
P_{1.4} \approx 1.3\times 10^{24} ~\left(\dfrac{M_{500}}{4\times 10^{14} ~ \mathrm{M}_{\odot}}\right)^{2.8} ~(\mathrm{W~Hz}^{-1}).
\end{equation}
Apart from the considerable scatter in this scaling relation, this also neglects all possible evolution of cluster properties relevant for the radio \syn emission. For SZ we used the self-similar redshift dependence $E(z)^{8/3}$ (Eq.\ref{eq:gnfw}) for scaling. Thus the above scaling law is clearly an approximation, but even for fixed luminosities for both \syn and the SZ signal, their observed flux ratio will scale as $(1+z)^4$ (the ratio $D_L^2/D_A^2$), on top of which there will be the strong $K$-correction term due to the steeply falling \syn spectrum. Thus, a strong $z$-increase in the flux contamination will be inevitable also with a more precise \syn power scaling.

\begin{figure}[ht]
%\centering
\includegraphics[width=\columnwidth]{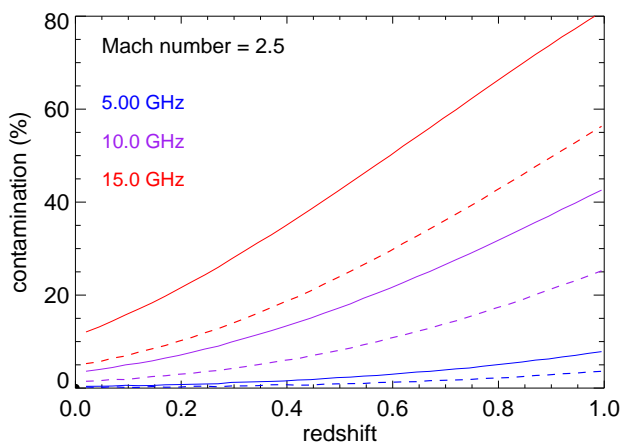}
\caption{SZ contamination in measured radio fluxes with increasing redshifts, for two different cluster masses. The solid lines are for a cluster with mass $M_{500} = 4\times 10^{14} \mathrm{M}_{\odot}$, and the dashed lines are for a higher mass $M_{500} = 1\times 10^{15} \mathrm{M}_{\odot}$. Three different colors mark three observing frequencies, and we use a $\mach=2.5$ shock with relic LLS $=0.5$ Mpc, at 1 Mpc  cluster-centric distance. 
We make use of use an empirical relic power-to-cluster mass scaling from \citet{deGas14}, $P_{\nu} \propto M^{2.8}$, which is stronger than the cluster SZ signal scaling. Hence the contamination decreases with increasing cluster mass.}
\label{fig:zcontam}
\end{figure}

In Fig. \ref{fig:zcontam} we show examples of the SZ flux contamination increase with redshift for two sets of cluster masses: $M_{500} = 4\times 10^{14} \mathrm{M}_{\odot}$ and $1\times 10^{15} \mathrm{M}_{\odot}$, using a relic model with $\mach = 2.5$ shock at $r_{\mathrm{shock}} = 1$ Mpc. 
It is seen that for 15 GHz observation the measured relic fluxes can be less than half of its true value at $z=0.5$ and above. 
\rev{This is important for several currently known intermediate- to high-$z$ relics,}  
%e.g. those in MACS clusters around redshifts $0.3-0.5$ \citep{Bona12}.
%, e.g., the relics in MACS J1752.1$+$4440 (Bonafede et al. 2012). 
\rev{one such example is discussed} in Section \ref{sec:elgordo} for the El Gordo cluster.

\subsubsection{Effect of the viewing angle}
\label{sec:angle}

For most of our work we are assuming that radio relics are viewed edge-on, i.e., the cluster merger and shock propagation is happening in the plane of the sky. While this is a reasonable assumption for double radio relics, for many other single relic observations the viewing angle remains an open and important issue. Prominent examples are the relics in the Coma and A2256 clusters, discussed later. Here we briefly present a qualitative description of the effect of changing viewing angles in the framework of our lognormal emissivity model on top of a spherical shock.

\begin{figure}[ht]
%\centering
\includegraphics[width=\columnwidth]{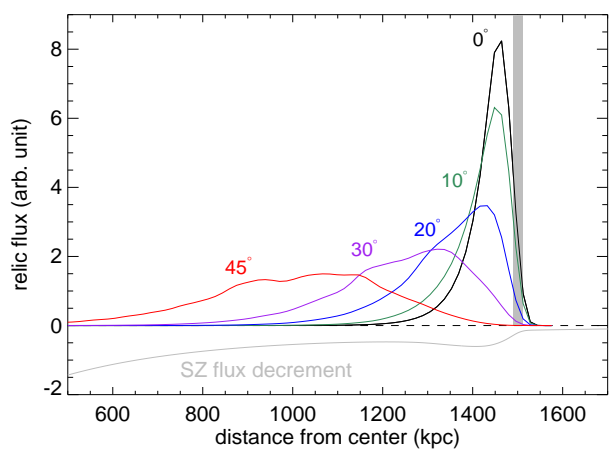}
\caption{How a radio relic with lognormal emissivity model will look like from different viewing angles. For this illustration we use a relic with dimension  500 kpc at a distance of 1.5 Mpc from a cluster center, situated at $z=0.2$, and simulate an observation with resolution $10\asec$. The grey vertical band shows the location of the shock front, and a negative SZ flux profile is included for comparison.}
\label{fig:rotation}
\end{figure}

Clearly, as the relic is viewed more and more obliquely (towards or away from the observer -- the cases are symmetric), its peak flux amplitude will drop while keeping the total flux conserved. If we assume a more extended spherical shock front for the SZ signal, this will also mean that the peak of the relic \syn flux will move away from the SZ signal jump, which is unaffected by rotation. This scenario is depicted in Fig. \ref{fig:rotation}, using the SZ and \syn signal modeling described earlier, but having an additional rotation perpendicular to the plane of the sky prior to projection. Detailed shapes of these profiles depend on numerous factors like relic opening angle and curvature, \rev{the flux profile along the length of the relic, etc.} What concerns us is the approximate shape with respect to the shock-front location. 
We can see that for even a mild rotational angle of $\sim 20^{\circ}$ the peak flux drops by roughly factor two. Additionally for the flux contamination issue, the bulk of the \syn emission shifts away from the SZ ``jump'' at the shock front, where the SZ signal is more uniformly negative. For a single-dish instrument we will get more flux contamination as the SZ signal is generally stronger near the cluster center, but an interferometer will generally filter out this uniform SZ signal and recover close to the true \syn flux. We 
%do not make simulations for interferometric observation for these cases, but 
shall present total-power flux contamination estimates for the Coma and A2256 relics in Sec. \ref{sec:indiv}.

\subsubsection{Effect of changing Mach numbers}
\label{sec:machdep}

Any change in the Mach number distribution at shocks will affect the relic \syn emission in a complicated way. 
In the simple scenario of single injection by a merger shock, we can assume that radio \syn power scales as $\mach^3$, as the particle energy injection is proportional to the kinetic power dissipated by the shock: $E_{\mathrm{kin}} = \mach^3 n_{\mathrm{u}} c_s^2 A/2$ (e.g. \citealt{Vaz14}; where  $n_{\mathrm{u}}$ is the upstream density, $c_s$ is the sound speed and $A$ is the shock surface area). In such a simplistic model the \rev{synchrotron}-to-SZ flux ratio will then scale linearly with $\mach$, since the SZ flux change is roughly proportional to $\mach^2$ at the shock front (Eq. \ref{eq:Pratio}). Thus we can expect a decrease in flux contamination in radio relics with higher Mach numbers.

A more sophisticated modeling can be obtained through the formalism by \citet{HB07}, that attempts to describe the stationary emission from a power-law population of accelerated electrons, coming from the Maxwellian tail of energetic enough electrons to enter diffusive shock acceleration. This model introduces an efficiency function that takes into account the particle spectra as a function of $\mach$ to compute the total 
injected energy into the electron population. A useful polynomial expression for this is given by the function $\eta({\mach})$ in \citet{Kang07} {\footnote{Compared to the original work by \citet{HB07}, the $\eta(\mach)$ function by \citet{Kang07} also takes into account the gas compression at the shock and must be applied to up-stream quantities. On the other hand, the efficiency given in \citet{HB07} is meant to be referred to down-stream quantities. In this work we follow the convention by \citet{Kang07} and therefore make use of up-stream gas quantities.}. This efficiency function has been derived for protons, and within the uncertainties related to the modeling of shocks it is appropriate also to use it to derive the electron acceleration efficiency, by rescaling for the electron-to-proton ratio, $\xi_{e/p}=0.05$ \citep{Kang13}. 
The characteristic feature of $\eta(\mach)$ is a very steep drop below $\mach \lesssim 3$, and a roughly constant value above $\mach \gtrsim 6$. With such a dependence we get a \rev{synchrotron}-to-SZ flux ratio as shown in Fig. \ref{fig:machdep} top panel. We also show the simplistic picture of $P_{\mathrm{sync.}} \propto \mach^3$ with dotted lines, where the flux ratio is linear.  
In the bottom panel of Fig. \ref{fig:machdep} we show the resulting spectrum for radio observation at cm-wavelengths. The increase in the low-frequency \syn flux amplitudes with increasing Mach numbers from the linear model is \rev{evident}.

\begin{figure}[ht]
%\centering
\includegraphics[width=\columnwidth, height=5.5cm]{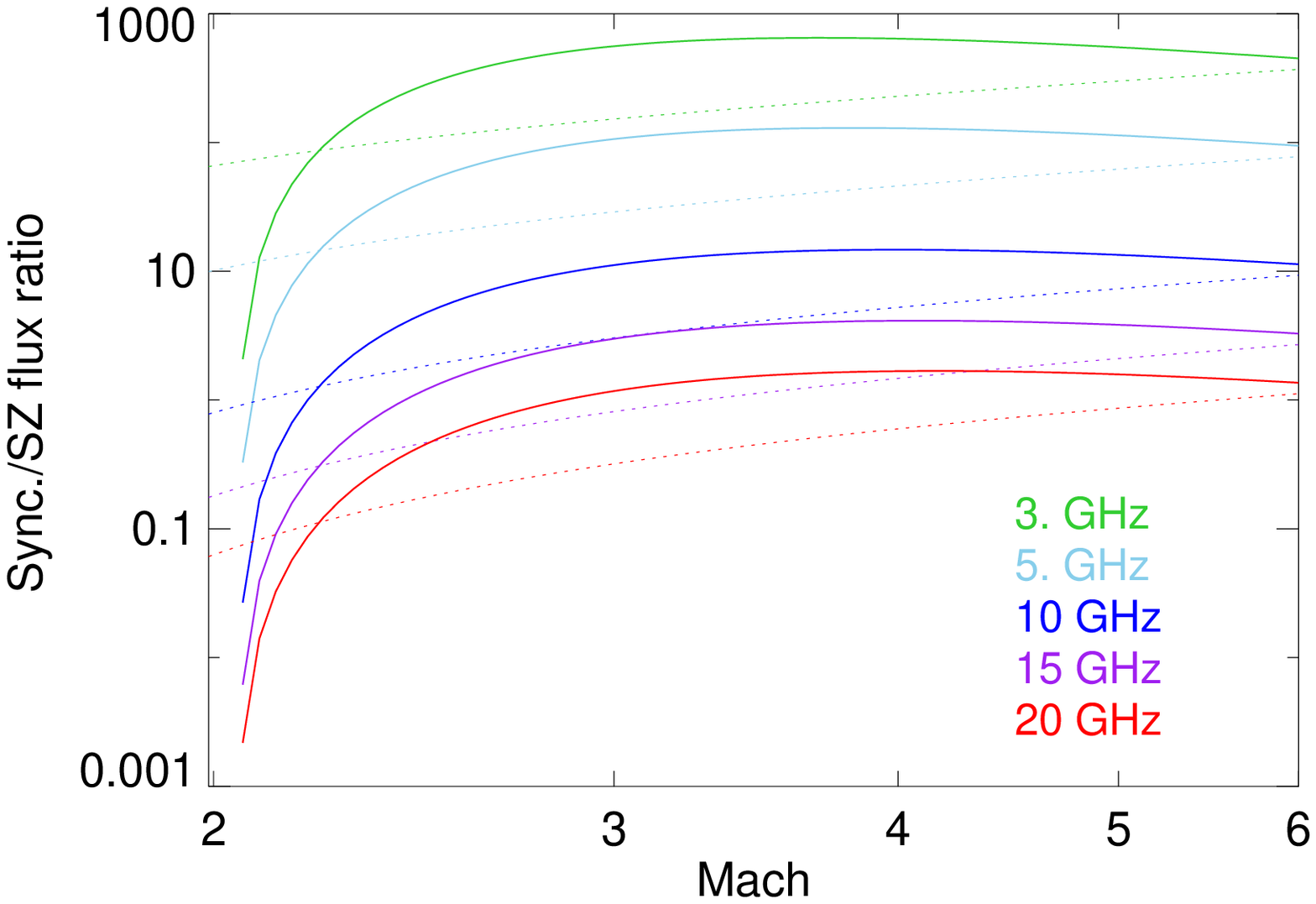}
\includegraphics[width=\columnwidth, height=6cm]{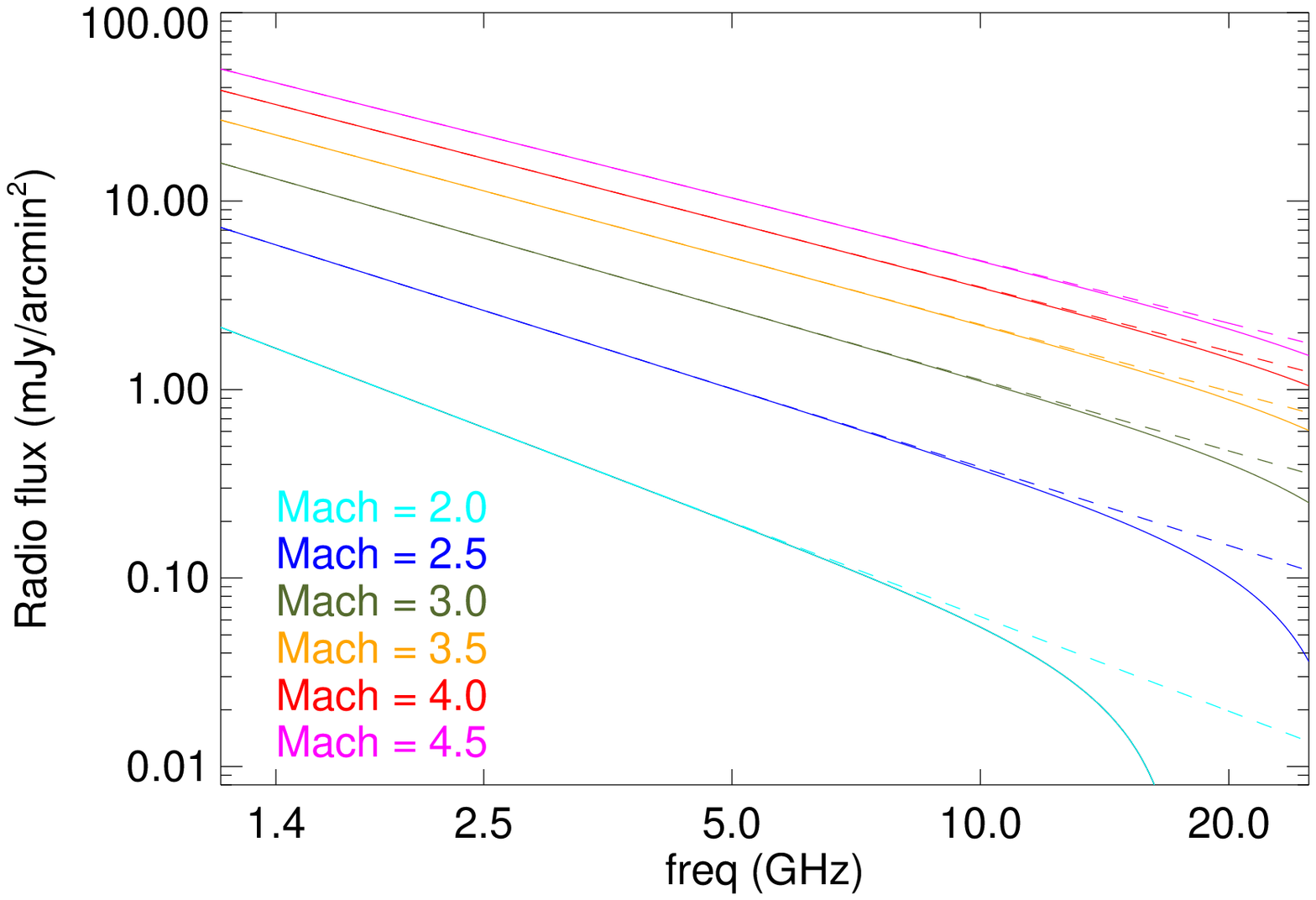}
\caption{Effect of a change in Mach number onto the measured radio spectrum. {\it Top panel:} \rev{Synchrotron}-to-SZ flux ratio as a function of the shock Mach number, seen at different observing frequencies. In the DSA-based model this ratio is varying rapidly only below $\mach \lesssim 2.5$, above which it asymptotes to a constant ratio which starts to decrease at high $\mach$-values. For comparison, results from a simple $\mach^3$ dependence of the radio \syn power is shown by the dotted lines, where the flux ratio increases linearly (see Sec. \ref{sec:theory}). 
{\it Bottom panel:} The observed radio spectra for different Mach number values, using the linear Mach dependence of the \rev{synchrotron}-to-SZ flux ratio. The dashed lines are uncontaminated \syn spectra that follow a single power-law throughout.}
\label{fig:machdep}
\end{figure}

The discussion above is forced to ignore several complications of the particle acceleration by merger shocks, which are yet to be solved (e.g. \citealt{Vaz15} and discussion therein).  
The presence of re-accelerated electrons can contribute to the \syn emission together with the freshly accelerated electrons, although this contribution is important only for $\mach \leq 3$ shocks \citep{Pinz13}. Amplification of the magnetic fields by cosmic rays-induced turbulence is expected to be important only for $\mach \geq 10$ \citep{Brug13} and therefore can be neglected. 
In Sec. \ref{sec:theory} we discuss some approximations to the DSA-based model to match its predictions to real radio relic data, where we neglect these secondary effects.
%to make rough estimates for the radio and SZ signals at relic shocks 

%%%%%%%%%%%%%%%%%%%%%%%%%%%%%%%%%%%%%%%%%%%%%%%%%%%%%%%%%%%%%%%%%%%%%%%%%%%%%%%%%%%%%%%
%\subsection{\textit{\textbf{Part II. Results for individual cluster relics}}}
\subsection{\textit{\textbf{Modeling results matching actual relic data}}}
\label{sec:indiv}

In this second part of our results we present detailed contamination modeling for some well-known radio relics, including the \sausage relic in CIZA J2242.8$+$5301 (Sec. \ref{sec:sausage}) and the NW relic in the El Gordo cluster (Sec. \ref{sec:elgordo}). Whenever possible we use actual measured values for the Mach numbers, and \rev{set the normalization of the GNFW pressure model to match the global SZ signals against  measured values.} %Most important contamination predictions are listed in Table \ref{onetable}. 
%A list of contamination predictions for these clusters at selected frequencies is shown in Table \ref{onetable}. 

\subsubsection{The \sausage relic in CIZA J2248.8$+$5301}
\label{sec:sausage}

We first present results for the flux contamination in the \sausage relic, as its 16 GHz flux measurements \citep{Str14} had been a main motivation for this work.    
Fit to the \sausage brightness profile with our lognormal model was discussed in Section \ref{sec:lognormal}, where we mentioned that the 1.7 Mpc length of the \sausage relic makes it difficult to obtain the sharp drop in radio \syn brightness seen in the post-shock end, because of projection. 
To avoid this difficulty, we simply model the {\it projected} brightness profile (at 610 MHz; \citealt{vanWe10}) directly with a lognormal model, which fit the flux-drop in the post-shock region very well, even though being less physically motivated. 
Another difficulty is the choice of Mach number for our modeling.   
From the \syn spectral index \citet{vanWe10} derived $\mach = 4.6$, whereas X-ray temperature measurements from the {\it Suzaku} satellite have yielded distinctly lower values in the range $2.7-3.2$ (\citealt{Aka13}, \citealt{Ogr13}, \citealt{Aka15}). 
We therefore present our results for three different Mach number possibilities ($\mach = 2.5, ~3.5, ~4.5$). The total spectral index of the relic is derived from the chosen Mach numbers (Eq. \ref{eq:alpha}) instead of using some fixed value\footnote{The low-frequency data for the \sausage relic (and also the \toothbrush relic discussed in Sec. \ref{sec:toothbrush}) are poorly fit by a single-power law, see e.g. \citet{Str15} for a discussion. This makes the choice of a unique Mach number to fit the spectrum difficult.}.
The other ingredients for computing the flux contamination are the relic length (LLS $= 1.7$ Mpc), distance from the cluster center ($r_{\mathrm{shock}} = 1.5$ Mpc), and the cluster mass.

For the \sausage relic cluster and other cluster results presented afterwards, we make an additional correction for the cluster mass based on the \plk all-sky measurements. The cluster masses are generally first estimated from their X-ray luminosities, but as all these relic clusters are merging systems, a large deviation from the mean scaling relations can be expected. The published all-sky Planck data \citep{Plk15} presents the possibility to directly calibrate the cluster SZ signal from $Y_{500}$ measurements. CIZA J2248.8$+$5301 is not in the published \plk catalog due to its proximity to the Galactic plane, and some other known relic clusters are absent either because they are low mass systems or too compact. To keep uniformity over all the cluster mass calibration results, we extract $Y_{500}$ by fitting a GNFW model to an $y$-map extracted from the \plk all-sky data (see \citealt{Er15} for details). For CIZA J2248.8$+$5301 we obtain $Y_{500} = (3.6 \pm 0.1)\times 10^{-3}$ arcmin$^2$, which is about factor 2 more than the GNFW model prediction based on the $L_X$-derived mass ($M_{500}^{\mathrm{X-ray}} = 5.5\times 10^{14}$ M$_{\odot}$). The upward-corrected mass is also very similar to the recent measurement for week lensing mass by \citet{Jee15}. 
Therefore, we scale the ambient pre-shock pressure north of the relic for a mass value in the GNFW model with $M_{500} = 8.5\times 10^{14}$ M$_{\odot}$, with the assumption that the boost in $Y_{500}$ due to the relic shock is small compared to the cluster's global SZ value.

The comparison between the \sausage relic data and our spectral computations is shown in Fig. \ref{fig:sausage} upper panel. The data points are interferometric measurements with comparable scales \citep{Str15}. We compute the \syn fluxes based on the 610 MHz data, but apply an additional scaling  such that the low-frequency part of the spectrum provides the best possible fit to all the low-frequency data ($\lesssim 2$ GHz). This way, our spectral prediction are not sensitive to any particular `pivot' flux value. 
For $\mach = 3.5$ we compute a contamination of less than 1\% at 5 GHz, and roughly 14\% at 16 GHz. This is within the $1\sigma$ statistical uncertainty in the AMI measurement quoted by \citet{Str14}. Above this frequency the flux contamination increases rapidly; for the same Mach number we have roughly 25\% flux contamination at 20 GHz, and 50\% contamination at 30 GHz. In other words, the measured radio flux is roughly half of \rev{the true synchrotron value} at 30 GHz.}  
The AMI 16 GHz measurement can be considered too low for our \rev{synchrotron}$+$SZ spectral model, particularly as the single power-law with $\mach=2.5$ provides a poor fit to the low-frequency part. This might be indicative on an intrinsic spectral steepening, however, any amplitude of such can only be robustly constrained after taking into account the SZ effect together with the relevant geometrical and observational parameters.

\begin{figure}[t]
%\centering
\includegraphics[width=\columnwidth, height=6cm]{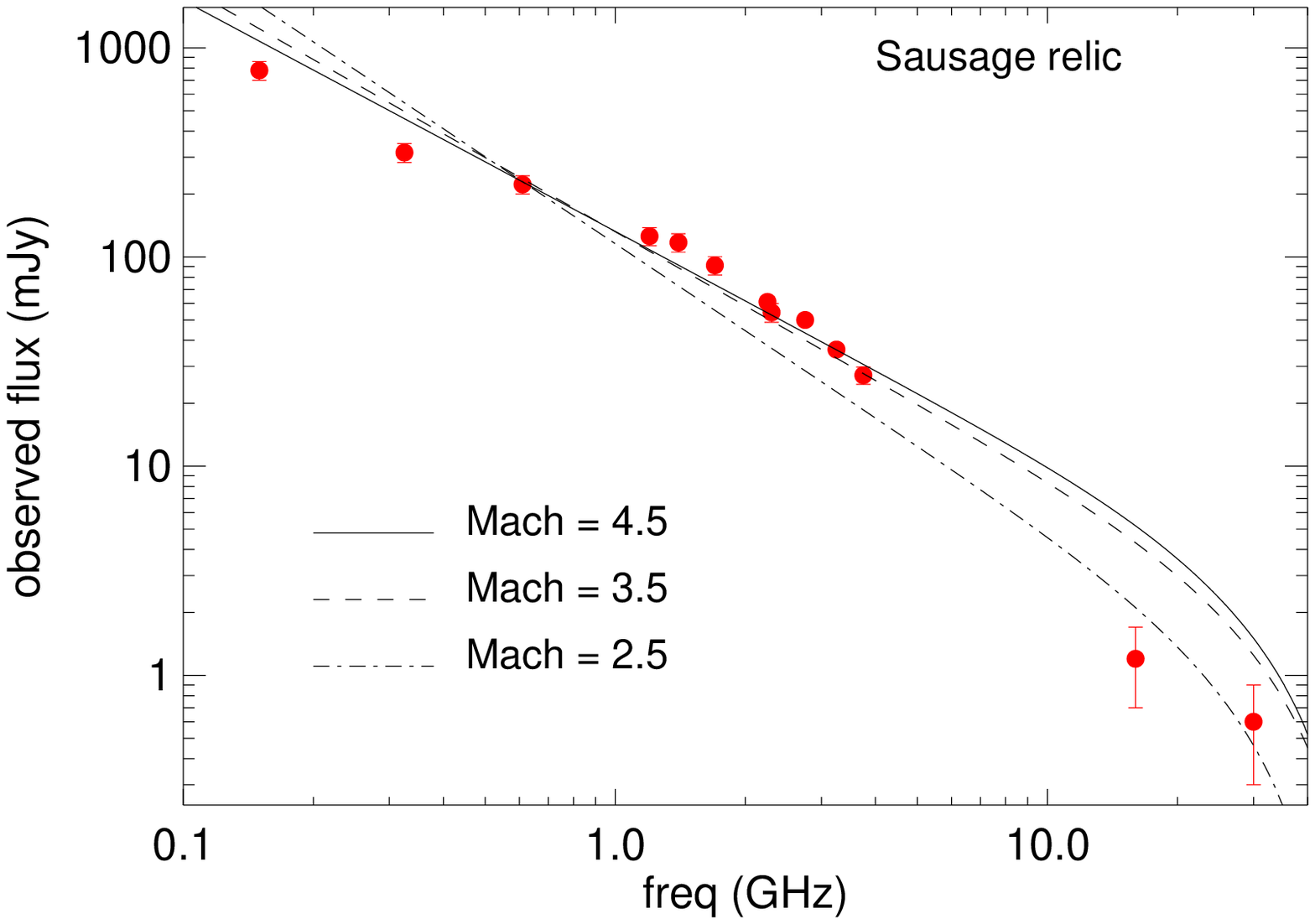}
\includegraphics[width=\columnwidth, height=6cm]{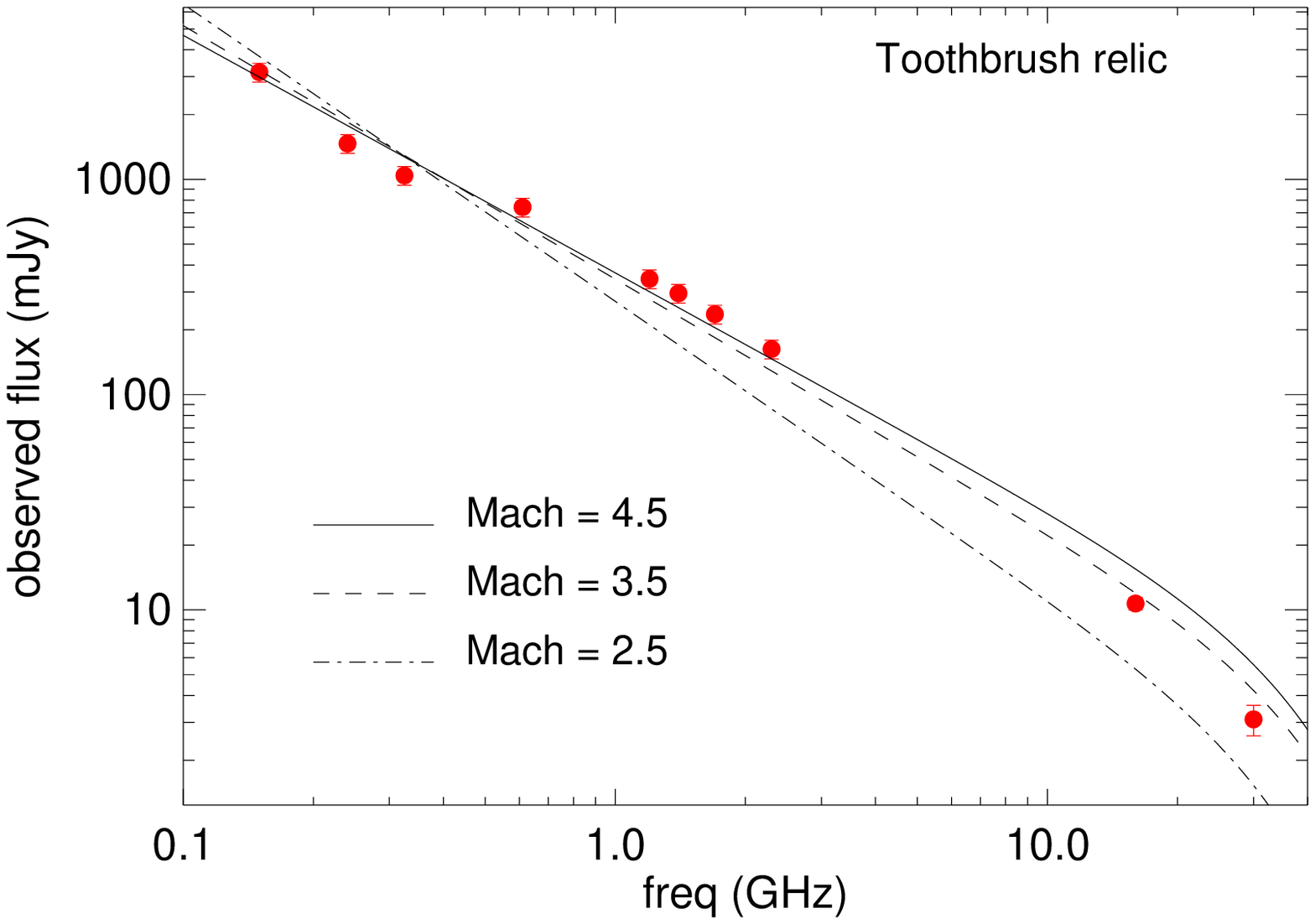}
\caption{Interferometric flux measurements of the \sausage relic in CIZA J2248.8$+$5301 \textit{(top panel)}, and the \toothbrush relic in 1RXS J0603.3$+$4214 \textit{(bottom panel)}, against our spectral predictions for three different shock Mach numbers. The data are taken from \citet{Str15}. The predicted spectra are normalized to have minimum $\chi^2$ computed from the low frequency ($\lesssim 2$ GHz) part. These model spectra correspond to an `ideal' measurement, and the results from actual interferometric imaging can differ. As discussed in Sec. \ref{sec:interferometry}, radio flux measured by interferometers in presence of SZ decrement  can be somewhat higher than shown here; the differences attributing to the relic angular size and interferometer configuration, measurement signal-to-noise, imaging procedure, etc. \rev{In addition, prediction for the SZ decrement can vary along the relic length depending on inhomogeneous ICM distribution, Mach number variation, etc.}}
\label{fig:sausage}
\end{figure}

At this level, our spectral modeling is not an accurate prediction for the specific interferometric measurements, in particular the critical high-frequency results from AMI and CARMA. This will require a detailed knowledge of the interferometric set-up and imaging procedures. The flux loss numbers shown here should be taken only as a guideline, not exact predictions for the real-life interferometric observation. As we showed in Sec. \ref{sec:interferometry}, interferometers will definitely compensate for some of the lost flux caused by the SZ decrement, but will also miss flux on large scales, particularly for nearby relics like {\it Sausage}. The ideal flux contamination numbers shown here are not a lower limit either, as the flux loss due to a lack of short baselines can go beyond the change caused by SZ alone. In Sec. \ref{sec:interferometry} we discussed the change of flux only in the transverse direction (for \sausage relic N-S direction), but the interferometric flux loss due to inadequate $uv$-sampling will affect the longitudinal (E-W) direction more, and the measured flux will reflect the combined effect. Following the discussion in \citet{Str15}, we do not use the single-dish `total power' spectra from their work, due to the possibility of point source contamination and wrong baselevel corrections.

From the predicted contamination percentage values listed in Table \ref{onetable}, one can note an interesting trend. For various Mach number values assumed, the flux contamination (i.e., the ratio of \syn to SZ flux) is roughly constant at any given frequency. In Section \ref{sec:machdep} we saw that in a DSA-based modeling the flux ratio reaches a constant value above $\mach \gtrsim 3$, but in the calculation for \sausage relic we have not assumed any specific model to predict the \syn flux. Instead, this behavior comes from computing the radio spectral index from the shock Mach number, which for our Mach-range of interest ($\mach \sim 2-4$) produce roughly $\alpha_{\mathrm{tot}} \propto \mach^{1/2}$. Thus when extrapolating \syn fluxes from a fixed measured value at low frequency, the log of \syn flux at a fixed higher frequency roughly scales as $\log(S_{\mathrm{sync.}}) \sim \mach^{1/2}$. The SZ flux boost, on the other hand, scales as $\mach^2$, i.e., $\log(S_{\mathrm{SZ}}) \sim 2\log(\mach)$. In the narrow Mach-range of interest these two functions produce roughly the same rate of change. Therefore, measuring the SZ flux contamination at one single frequency may not be a sensitive test for determining the shock Mach number. However, if the \syn spectral index can be robustly constrained using low frequency data, then measurement of the spectral steepening will readily point to the amplitude of the SZ effect, which in turn can be used for an {\it independent} derivation of the Mach number given some estimates for the pre-shock density and temperature. 
We will further explore this issue  in Section \ref{sec:disc}.

\begin{table*}[ht]
\begin{center}
\begin{tabular}{|r|c|c|c|c|c|c|}
\hline\hline
  &  3 GHz  &  5 GHz &  10 GHz  &  15 GHz  &   20 GHz  &  30 GHz  \\
  \hline
  &   &    &    &    &    &    \\  

Sausage relic (${\cal M}=2.5$) &  $<1$\%  &  $<1$\%  & 4\%   &  11\%  &  24\%  &  58\%  \\
(${\cal M}=3.5$) &  $<1$\%  &  $<1$\%  & 3\%   &  10\%  &  21\%  &  49\%  \\
(${\cal M}=4.5$) &  $<1$\%  &  $<1$\%  & 4\%   &  12\%  &  24\%  &  52\%  \\
 &   &    &    &    &    &    \\  
 
Toothbrush relic  (${\cal M}=3.5$) &  $<1$\% & $<1$\%   & 3\%   & 9\%   &  18\%   &   43\%  \\
(${\cal M}=4.5$) &  $<1$\%  &  $<1$\%  & 3\%   &  10\%  &  20\%  &  46\%  \\

 &   &    &    &    &    &    \\  

El Gordo relic  (${\cal M}=2.5$) &  $<1$\% & 3\%   & 23\%   & 53\%   &  81\%   &   $>100$\%  \\

 &   &    &    &    &    &    \\  

A2256 relic (${\cal M}=2.0$) &  1\% &  3\%   &  28\%  &  66\%   &  96\%   &  $>100$\%  \\

  &   &    &    &    &    &    \\  

Coma relic  (${\cal M}=2.2$) &  $<1$\% &  $<1$\%   &  6\%  &  21\%   &  41\%   &  82\%  \\

  &   &    &    &    &    &    \\  
\hline\hline
  
\end{tabular}
\end{center}
\caption{SZ flux contamination predictions, in percentage, for some well-known radio relics at different frequencies. We define the contamination value as $C (\%) \equiv 100\times (\mathrm{true~ flux} - \mathrm{measured~ flux})/\mathrm{true~ flux}$. The flux correction factor, $F$, which one should multiply with to get the true \syn flux values, is therefore $F = 100/(100-C)$. 
From this definition, $C$ is equal to the flux ratio between SZ and \syn (see Sec. \ref{sec:theory}): $C \simeq 100 \times (S_{\nu}^{\mathrm{SZ}}/S_{\nu}^{\mathrm{sync.}})$, {\it provided} all the fluxes are measured over the same area. In practice, we calculate the observed radio flux by integrating over the area where relic emission is positive, approximating an ideal total-power measurement. 
In the table a contamination value more than 100\% means that the \syn flux is completely overshadowed by the negative SZ signal. }
\label{onetable}
\end{table*}

\subsubsection{The \toothbrush relic in 1RXS J0603.3$+$4214}
\label{sec:toothbrush}
The famous \toothbrush relic is the other example where a spectral steepening has been observed at high frequencies \citep{Str15}, and the modeling of its SZ contamination is almost identical in procedure as in the \sausage relic. The $\sim 2$ Mpc linear shape of the relic is peculiar, but following \citet{Bru12} we assume that it is also caused by a merger shock. The Mach number determined from radio and X-ray data are similarly discrepant as in {\it Sausage}: \citet{vanWe12} obtained the radio \syn spectrum and determined a high $\mach = 4.6$ based on DSA theory, whereas {\it Suzaku} X-ray analysis by \citet{Ita15} provide a value of only $\mach = 1.5$. The cluster X-ray temperature of 7.3 keV quoted by \citet{Ita15} gives an X-ray based mass of roughly $M_{500}^{\mathrm{X-ray}} = 6.6 \times 10^{14}$ M$_{\odot}$, and a \plk derived $Y_{500}$ value of $(4.1\pm 0.1)\times 10^{-3}$ arcmin$^2$ suggests roughly 70\% more mass for the pressure calculation. The assumed relic geometry is $LLS=1.9$ Mpc and $r_{\mathrm{shock}} = 1.08$ Mpc, from \citet{vanWe12}.

The spectra for the \toothbrush relic, shown in the bottom panel of Fig. \ref{fig:sausage}, provides a much better fit to the high-frequency measurements of \citet{Str15}, as compared to the \sausage relic. The contamination values are very similar, and the modified spectra give support to  a high value of Mach number ($\mach \sim 3.5$) than those suggested from X-ray measurements (although, as in the \sausage spectrum, the low-frequency data can not be fit satisfactorily by a single power-law spectrum).

\subsubsection{The NW relic in El Gordo (ACT-CL J0102$-$4915)}
\label{sec:elgordo}

The two El Gordo cluster relics \citep{Lind14} present an exciting case for computing the SZ flux contamination at cm-wavelengths. At $z=0.87$ this cluster hosts both the highest redshift radio relics and radio halo known to-date, and we already discussed the case for its radio halo in Section \ref{sec:szhalo}. Here we present the SZ flux contamination estimation for its more prominent of the two relics, the NW relic. Its highest frequency measurement is at 2.1 GHz with ATCA, and some evidence for spectral steepening is found from the inter-bandpass measurement ($\alpha_{0.6}^{2.1} = 1.2\pm 0.1$ and $\alpha_{1.6}^{2.6} = 2.0\pm 0.2$; see \citealt{Lind14}). We check whether the SZ flux contamination can be effective already at this very low frequency.

For the El Gordo NW relic our lognormal emissivity model provides an excellent fit to the published radio \syn brightness profile (Fig. \ref{fig:lognormfit}), supporting an edge-on relic. The GNFW-model based global SZ prediction for the cluster, with X-ray derived mass $M_{500} = 11.7\times 10^{14}$ M$_{\odot}$ \citep{Men12},  is consistent with the ACT and SPT published values  (\citealt{Marri11}, \citealt{Willi11}), so no additional SZ flux correction is needed. There is only a radio spectral-index based measurement of the Mach number, $\mach = 2.5^{+0.7}_{-0.3}$ \citep{Lind14}, which we adopt here. The results do not support any significant flux contamination at the published ATCA band: at 3 GHz the flux contamination is less than 1\%. 
At 5 GHz the flux contamination is 3\%, but thereafter it increases rapidly: at 15 GHz only half of the true \syn flux can be observed, and at 30 GHz the \syn flux is completely swamped by the negative SZ signal (Table \ref{onetable}). This shows that measurements along the line of \sausage relic will be completely erroneous for the El Gordo cluster already at 10 GHz if the SZ effect is not taken into account, even though the weak spectral steepening suggested from the current low-frequency data appears safe from SZ contamination.

\subsubsection{A sample of double (edge-on) relics}

We present SZ contamination result predictions for a near-complete sample of double relic clusters, as compiled by \citet{deGas14} and \citet{Vaz15}. We can reasonably assume all these relics are viewed edge-on. Reported radio spectral indices are used to derive the shock Mach numbers. To compute the SZ fluxes, first cluster masses are derived from the soft-band X-ray luminosities given in the above publications, 
and then the resulting SZ signals are corrected by the $Y_{500}$ values extracted uniformly from the $y$-maps derived from \plk data (see Sec. \ref{sec:sausage}). 
All other relevant information, like the relic LLS and cluster-centric distances, and the 1.4 GHz radio power, are available for this compilation.

\begin{figure}[ht]
\centering
\includegraphics[width=\columnwidth]{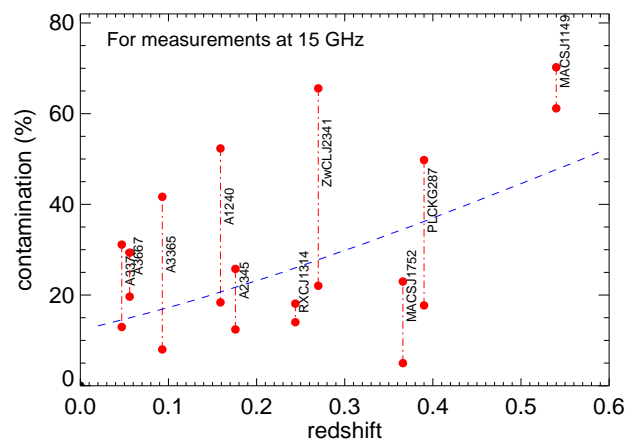}
\caption{Contamination predictions for a set of double relics in ten galaxy clusters, for whom the edge-on geometry can be applied with reasonable confidence. The relic dimensions and flux measurements are compiled by de Gasperin et al. (2014) and Vazza et al. (2015); the SZ predictions are calibrated against actual $Y_{500}$ values derived from \plk data. 
The blue dashed line is the mean contamination prediction from our modeling (see Section \ref{sec:zdep}), using the mean cluster mass and shock parameters for this sample. 
Red dot-dashed lines connect the relic pairs for easier identification. }
\label{fig:sample}
\end{figure}

In Fig. \ref{fig:sample} the contamination predictions at 15 GHz are shown for this double relic sample. The mean contamination at this frequency is in the range $10-50\%$, increasing with redshift, which we show by the blue-dashed line. This curve corresponds to the mean mass and Mach number ($\bar{M}_{500} \sim 5.5\times 10^{14}$ M$_{\odot}$, $\bar{\mach} \sim 2.5$) for the sample. 
The actual cluster-specific values exhibit a large scatter around this mean curve, and furthermore, we see that 
 the SZ contamination for two relics in the same cluster can differ by a factor of several. This is due both for the difference in their observed radio powers and also their cluster-centric distances, which result in very different ambient (pre-shock) pressures.
Combined with the difficulty of determining accurately the shock Mach number from radio data, this scatter in the SZ contamination can make it very uncertain to use high-frequency relic measurements to constrain the general properties of shock acceleration or electron ageing.

\subsubsection{Radio relic in A2256}
\label{sec:a2256}

The radio relic in A2256 is the only case where an attempt has been made to estimate the SZ flux contamination, by \citet{Tra15}. 
In that work, 10 GHz data from Effelsberg single-dish receiver were presented, and using a simple order-of-magnitude calculation 
%{we} 
\last{the SZ flux contamination was estimated to be} of the order 25\% at that frequency. However, it was argued that this contamination is an upper limit of the effect, and in any case, the concern of the paper was with the exceedingly flat \syn spectrum that is in tension with simple DSA-based stationary shock models. Given the challenges faced by the standard relic models to explain the A2256 radio emission (see e.g., \citealt{ClEns06}, \citealt{vanWe12}, \citealt{Tra15}), we do not attempt to explain the emission scenario, but rather use the a-priori assumption that the emission is indeed a shock-induced radio relic, and make a realistic estimate for the SZ contamination based on our model.

\begin{figure}[t]
%\centering
\includegraphics[width=\columnwidth, height=6cm]{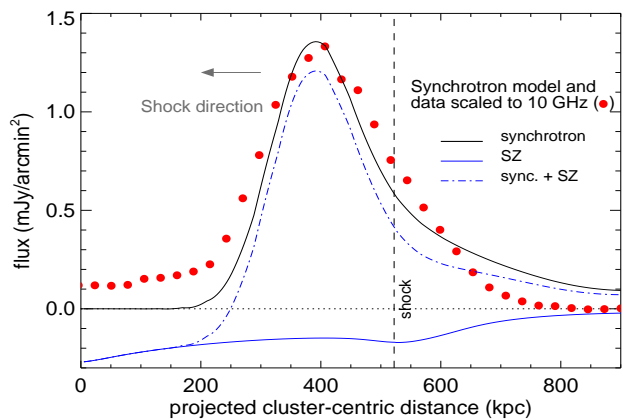}
\caption{A schematic model for the \syn and SZ flux profiles for the A2256 relic at 10 GHz. The geometry assumes a $\mach = 2$ shock propagating inwards at roughly 500 kpc distance from the cluster center, and viewed at an angle of $40^{\circ}$. The black solid line is the \syn flux profile, and the blue dot-dashed line is the SZ profile. The red points are the measured 1.4 GHz flux density values from \citet{ClEns06}, scaled to the \rev{considered} frequency.}
\label{fig:a2256}
\end{figure}

The modeling of the A2256 \syn emission profile poses additional problem with our adopted lognormal emissivity model: it cannot be fitted very well with the standard outwardly propagating shock that we have considered so far. In addition there is the strong likelihood that the relic is viewed at a large angle. We find acceptable fit (Fig. \ref{fig:a2256}) by allowing the shock to move {\it inwards}, when the relic is viewed at approximately $40^{\circ}$. This scenario is almost identical to the one presented by \citet{ClEns06} in their Fig. 11a.  The peak of the relic emission is \rev{at a projected distance} of 400 kpc from the cluster center \citep{vanWe12}, which means the pre-shock material is actually 520 kpc away, and hence the ambient SZ signal lower by roughly factor 1.6. In Fig. \ref{fig:a2256} we show the expected fluxes at 10 GHz, where the \syn is scaled down by a \rev{spectral slope $\alpha_{\mathrm{tot}} = 0.81$ \citep{vanWe12}}. The red data points correspond to the scaled-down version of the actual flux profile measured by \citet{ClEns06}. We assume the shock Mach number to be $\mach = 2$, for simplicity and for comparison with the estimate in \citet{Tra15}. One consequence of the large viewing angle and our assumption of a full-spherical shock for the SZ signal is that the \syn emission appears strongly shifted from the SZ shock feature, as is seen in Fig. \ref{fig:a2256}. The relic falls into the more `uniform' SZ decrement area, and as such interferometric measurements might be able to recover almost all the lost flux, but the published  $5-10$ GHz measurements of \citet{Tra15} are made with a single-dish (Effelsberg 100-m) and should bear the full effect of the large-scale SZ decrement.

After adjusting the cluster pressure profile to the {\it Planck} derived $Y_{500}$ value (equivalent mass $M_{500} = 7.9\times 10^{14}$ M$_{\odot}$), and assuming the shock and relic geometry as described above, we predict a radio flux contamination of roughly 1\% at 3 GHz, and 3\% at 5 GHz. These numbers are smaller than the current measurement uncertainties. At 10 GHz the flux contamination is more significant, roughly 28\%. This also turns out to be nearly identical to the \last{rough estimate made in \citet{Tra15}}, calculating an SZ flux of $\sim -20$ mJy compared to the measured radio flux $61.7$ mJy at 10.45 GHz. Nevertheless, contrary to the suggestion given in \citet{Tra15}, our new estimates should not be considered as an upper limit for the SZ effect. Several factors can result in a stronger SZ contamination, like closer distance to the cluster center, a higher shock Mach number, or integrating over some negative signal in a single-dish observation. From our current modeling and idealized flux measurement set-up, we predict that the \syn signal will be completely dwarfed by the SZ decrement above 20 GHz for the A2256 relic.

The contamination values for the A2256 relic are much stronger than the other low-$z$ relics we have considered so far (see Table \ref{onetable}). One reason is the low radio luminosity for this relic compared to both \sausage and \toothbrush relics given their respective cluster masses: the total 1.4 GHz rest-frame power for A2256 is $P_{1.4} = 3.6 \times 10^{24}$ W/Hz \citep{ClEns06}, about factor 2.5 lower than the mean relic power scaling of \citet{deGas14}, whereas both \sausage and \toothbrush relics are highly over-luminous. 
It is also possible that we over-estimate the SZ due to projection from our full-spherical shock model, when the relic is viewed at a large angle. 
However, in such a case the cluster-wide SZ signal will be the main contributor for the flux decrement (Fig. \ref{fig:a2256}), as the relic appears closer to the cluster center.
%Yet another possibility, related to the low apparent luminosity, is that the relic is viewed at large angle, and therefore appears closer to the cluster center (where the cluster-wide SZ decrement is stronger). 
%In our modeling we have assumed a full-spherical shock for projecting the SZ signal, and computed the resulting contamination fraction by taking the ratio of observed positive flux against the true radio flux. 
%%An interferometric imaging will see a lower contamination value as the radio signal falls on top of the `smooth' global SZ signal of the cluster. 
Similar considerations also affect the Coma relic measurement as discussed below.

\subsubsection{The Coma relic}
\label{sec:coma}

The Coma cluster hosts the other well-known radio relic in the local universe, and in \citet{Er15} we have presented the first measurement of a radio relic shock Mach number through the SZ signal variation. The results in \citet{Er15} were based on a cylindrical shock model, and for consistency with the other results in this paper we fit the {\it Planck} derived Coma SZ profile with a spherical shock model. We also use the latest 2015 data release of {\it Planck} for the $y$-map extraction. The results is a shock with $\mach = 2.2\pm 0.3$, which is consistent with the \citet{Er15} result. The shock radius is fixed at 2.1 Mpc from Coma cluster center (75$^{\prime}$). Unlike A2256, the current multi-wavelength data clearly supports an outwards-moving shock for the Coma relic (see discussion in \citealt{Er15}), even though several earlier works have argued for a different scenario (e.g., \citealt{Fer06}, \citealt{Brown11}). Modeling the relic \syn emission with our lognormal model also lends support to an outwards moving shock, where the relic is seen at an angle of roughly 30$^{\circ}$ (see Fig. \ref{fig:profiles}). Interestingly, this is very close to the viewing angle of $\sim 37^{\circ}$ originally proposed by \citet{Enss98}.

We use the total relic flux and spectral index from \citet{Fer12}, and the cluster-wide GNFW pressure fit result obtained in \citet{Er15}. The contamination estimates are listed in Table \ref{onetable}. Up to 10 GHz the effect of SZ contamination is small, but at 20 GHz the contamination will be approximately 40\%, and at 30 GHz 80\%. Similar to A2256, these contamination values are also high for a local universe cluster, likely for the same reasons discussed earlier (in particular, the radio luminosity is order of magnitude lower than the mean relation of \citealt{deGas14}). Present and future single-dish observations of the Coma relic should exercise particular care in modeling its total flux above 10 GHz.

%%%%%%%%%%%%%%%%%%%%%%%%%%%%%%%%%%%%%%%%%%%%%%%%%%%%%%%%%%%%%%%%%%%%%%%%%%%%%%%%%%%%%%%
\section{Discussion}
\label{sec:disc}
%%%%%%%%%%%%%%%%%%%%%%%%%%%%%%%%%%%%%%%%%%%%%%%%%%%%%%%%%%%%%%%%%%%%%%%%%%%%%%%%%%%%%%%

We have shown that the SZ flux contamination is something that can not be ignored when interpreting the current and future cm-wavelength data for cluster radio relics, but  can we learn something {\it new} from this? In this section we speculate on the implications of the thermal vs non-thermal measurements at cm-wavelengths, and derive a simple analytical expression for the \rev{synchrotron}-to-SZ flux ratio. To demonstrate the practical advantages that might be gained from measuring this ratio, we propose  a possible method to constrain the relic magnetic field or the shock acceleration efficiency.  

\subsection{Thermal modification of the non-thermal signal}
\label{sec:nontherm}

A significant SZ decrement at cm-wavelengths is an unavoidable result of the widely assumed shock-relic connection. 
%The SZ signal change must originate from roughly the  same scales responsible for the observed radio emission. 
The only way to avoid SZ contamination is to assume that the gas \rev{pressure} distribution \rev{across the relic location} is much smoother than the radio \syn emission (so that \last{the SZ signal} can be almost entirely filtered out by radio-interferometric observations). However, this would force a complete reconsideration of the shock scenario necessary for the electron acceleration and the emergence of relics.  
The uncertainties in the actual amplitude of the SZ decrement only depends on ``simple'' physical ingredients, such as the unknown gas pressure, the Mach number, the exact profile of the shock and the geometry of the system. These factors make our estimate uncertain within a factor $\sim 2$ in most cases. Given this uncertainty, the fact that the estimated SZ decrement is of the order of the observed steepening of high frequency radio spectra of relics can be regarded as one of the best evidence so far that relics are actually connected to shocks, as the two signals (\syn emission and SZ decrement) are nearly equally sampled by radio observations.

Once we assume that the observed spectral steepening is solely due to the SZ effect, then the spectral curvature provides a straightforward way to measure the SZ effect itself, and model the shock Mach number (if the spectrum at the low-frequency end is modeled by a single power-law). 
This `derived' SZ amplitude can be compared against independent SZ measurements at millimeter/sub-millimeter wavelengths, or from combining X-ray results. If a spectral steepening is found even after correcting for the SZ flux bias (as is apparently happening in the NW relic of El Gordo), then a case can be made for non-standard scenarios in the shock acceleration model. An additional test can be for the shock Mach number itself: the current thermal (X-ray) and non-thermal (\rev{synchrotron}) Mach number estimates are often in tension, whereas we have assumed the same shock Mach number for SZ and \syn flux modeling throughout. If observations of the low-frequency \syn spectral slope and the high-frequency spectral steepening can both be explained by the same Mach number, then that will be a strong indicator for the true Mach number of the underlying merger shock.

\subsection{\rev{Analytical model for the synchrotron-to-SZ flux ratio}}
\label{sec:theory}

We now wish to derive a simple analytical expression for the \rev{synchrotron}-to-SZ flux ratio, 
%\rev{hereafter called simply as the radio-to-SZ flux ratio},  
that can provide a quantitative measure for the spectral steepening at cm-wavelengths. This requires connecting the properties of the cosmic ray electrons to those in the thermal gas, which brings in heavily model-dependent assumptions. 
For our illustration, we use the DSA-based model for relic radio \syn emission by \citet{HB07}, in the simplified form as given by \citet{Vaz15}: 
\begin{equation}
\begin{split}
P_{\nu}^{\mathrm{~sync.}} \approx ~& 3.5\times 10^{26}~ \mathrm{W/Hz}~ 
	\left(\dfrac{L^2}{1~ \mathrm{Mpc}^2}\right)~ 
	\left(\dfrac{n_{\mathrm{u}}}{10^{-4} \mathrm{cm}^{-3}}\right)~ \left(\dfrac{T_{\mathrm{u}}}{1 \mathrm{~keV}}\right)^{3/2} \\
	&~~ \times \left(\dfrac{\nu}{1.4~ \mathrm{GHz}}\right)^{-\delta/2} 
	\dfrac{B_{\mathrm{relic}}^{1+\delta/2}}{B_{\mathrm{CMB}}^2 + B_{\mathrm{relic}}^2}~ 
	\left(\dfrac{\xi_{e/p}}{0.05}\right)~ \eta(\mach). 
\end{split}
\label{eq:rmodel}	
\end{equation}
In this formulation of the total relic power, $L$ is the observed relic length (hence approximating the relic surface area as $A \approx L^2$), $n_{\mathrm{u}}$ and $T_{\mathrm{u}}$ are the upstream (pre-shock) electron density and temperature, $\xi_{e/p}$ is the electron-to-proton ratio, and $\eta(\mach)$ is the shock acceleration efficiency. The spectral index and the magnetic field carry Mach number dependence as well, through the electron momentum power-law index $\delta$ (cf. Eq. \ref{eq:alpha}).  
%From Eq. \ref{eq:alpha} it follows $\delta = 2(\mach^2+1)/(\mach^2-1)$. 
$B_{\mathrm{relic}}$ is the relic magnetic field in $\mu$G, and $B_{\mathrm{CMB}} = 3.24 (1+z)^2 ~\mu$G. 

The simplified expression of Eq. \ref{eq:rmodel} can be expected to match real observations only within an order of magnitude: the main theoretical uncertainty comes from the electron acceleration efficiency, given by $\xi_{e/p} \cdot \eta(\mach)$. We described its behavior in the DSA-based model of \citet{HB07} in Section \ref{sec:machdep}, referring to the fitting formula in \citet{Kang07} for the case without pre-existing CRe component when $\mach > 2$. 
However, there is much theoretical uncertainty in this term, and as \citet{Kang12} have shown, several combinations of pre-shock cosmic rays and injection efficiency can explain the same observed relic emission. 
The fitting formula of \citet{Kang07} can be approximated near $\mach \sim 3$ by $\eta(\mach) \approx 0.004\mach^3$, however, using this approximation the \syn power for the \sausage relic is over-predicted by a factor of 13 (against the measured value of $P_{1.4} = 1.5\times 10^{25}$ W/Hz; \citealt{deGas14}). In reality this discrepancy can be due to several other unknowns, like the upstream density, CR e/p ratio, or the relic magnetic field; but we make the assumption that this comes mainly from the efficiency term $\eta(\mach)$, and fix its normalization as $\eta(\mach) \approx 3\times 10^{-4} \mach^3$ near $\mach \sim 3$. With this approximation the Eq. \ref{eq:rmodel} matches real observations more closely.

After this `re-calibration' of the radio \syn power, we can write the {\it total} monochromatic flux observed from a radio relic, assuming standard isotropic flux distribution, as 
\begin{equation}
\begin{split}
S_{\nu}^{\mathrm{sync.}} & = (1+z)^{-(\delta/2)}~ P_{\nu}^{\mathrm{~sync.}} / 4\pi D_L^2 \\
	& \approx 24 ~\mathrm{mJy}~ \left(\dfrac{\mach}{3}\right)^3~ \left(\dfrac{\xi_{e/p}}{0.05}\right)~ 
	\left(\dfrac{L^2}{1~ \mathrm{Mpc}^2}\right)~\dfrac{B_{\mathrm{relic}}^{1+\delta/2}}{B_{\mathrm{CMB}}^2 + B_{\mathrm{relic}}^2} \\
	&~~ \times \left(\dfrac{n_{\mathrm{u}}}{10^{-4} \mathrm{cm}^{-3}}\right)~ 
	\left(\dfrac{T_{\mathrm{u}}}{1 \mathrm{~keV}}\right)^{3/2}~ \left(\dfrac{D_L}{10^3~ \mathrm{Mpc}}\right)^{-2} \\
	&~~ \times (1+z)^{-\delta/2} ~\left(\dfrac{\nu}{1.4~ \mathrm{GHz}}\right)^{-\delta/2} 
\end{split}
\label{eq:Snuradio}
\end{equation}
%In the above flux formula we added a  $K$-correction term: $K(z) = (1+z)^{-\delta/2}$, using the integrated relic spectral index $\alpha_{\mathrm{tot}} = \delta/2$. 
\last{In the above flux formula $(1+z)^{-\delta/2}$ represents the $K$-correction due to the falling synchrotron spectrum with index $\alpha_{\mathrm{tot}} = \delta/2$, ignoring an additional $(1+z)$ term for the bandwidth.} 
$D_L$ is the luminosity distance, and in the second step we assumed a cluster at $z=0.2$ such that $D_L \approx 10^3$ Mpc. Henceforth we are also  using the approximation $\eta(\mach) \approx 3\times 10^{-4}\mach^3$, which greatly simplifies the Mach dependence in our formulas but is only valid for a narrow range around $\mach\sim 3$.

The SZ ``flux'' is not isotropic, it depends on the line-of-sight projection of the pressure, and therefore relies on the shock geometry at the location of the relic. For an edge-on relic we can consider only the shock-boosted pressure at the relic location, ignoring the main cluster's SZ contribution. To simplify the projection let us also consider plane-parallel geometry, ignoring curvature. The line-of-sight depth is the same as the relic length in the plane of the sky, $L$. Then from Eq. \ref{eq:y} the Compton-$y$ parameter just behind the shock front would be
\begin{equation}
y \approx 3.6\times 10^{-6} \left(\dfrac{n_{\mathrm{u}} T_{\mathrm{u}}}{10^{-4}~ \mathrm{keV~ cm}^{-3}}\right)~ \left(\dfrac{L}{1~ \mathrm{Mpc}}\right)~ 
	\left(\dfrac{\mach}{3}\right)^2.
\label{eq:realy}
\end{equation}
In keeping with the \syn modeling we have taken the relic LLS to be 1 Mpc and shock Mach number $\sim 3$. 
We also approximated the shock-boosted pressure in Eq. \ref{eq:Pratio} as $P_\mathrm{d}/P_\mathrm{u} \approx \mach^2$. 
This $y$-value translates into a brightness (or Rayleigh-Jeans) temperature, at frequencies $\lesssim 30$ GHz, of roughly  $\Delta T_{\mathrm{RJ}} \approx -20~ \mu$K, with the same shock parameters as in Eq. \ref{eq:realy} above. 

To compute the total SZ flux we need to integrate $y$ over the visible relic area.  
Let us take the mean relic width to be ${\cal W} \approx 100$ kpc, such that the angular size of the relic is $\Omega_{\mathrm{relic}} \approx L{\cal W}/D_A^2$ steradians, where $D_A$ is the angular diameter distance. Therefore, applying Eq. \ref{eq:szflux} we obtain the following expression for the {\it total} SZ flux decrement 
\begin{equation}
\begin{split}
S_{\nu}^{\mathrm{SZ}} \approx \mathcorrect{-0.26} & ~\mu\mathrm{Jy}~ \left(\dfrac{D_A}{700~ \mathrm{Mpc}}\right)^{-2} 
	\left(\dfrac{L}{1~ \mathrm{Mpc}}\right)^2 \left(\dfrac{\cal W}{100~ \mathrm{kpc}}\right) \\
	& \times \left(\dfrac{n_{\mathrm{u}} T_{\mathrm{u}}}{10^{-4}~ \mathrm{keV~ cm}^{-3}}\right)~ 
	\left(\dfrac{\mach}{3}\right)^2 \left(\dfrac{\nu}{1.4~ \mathrm{GHz}}\right)^2 
\end{split}	
\label{eq:Snusz}
\end{equation}
The SZ flux is proportional to $\nu_{\mathrm{obs}}^2$ in the RJ-limit, and we have taken $D_A = 700$ Mpc for a $z\sim 0.2$ cluster in accordance with the derivation above. We see that the SZ signal has the same $L^2$ dependence as in the \syn, with an additional width term ${\cal W}$ that needs to be determined observationally.

%Combining Eqns. \ref{eq:Snuradio} and \ref{eq:Snusz}, we get the radio-to-SZ flux ratio as a function of frequency: 
The \rev{synchrotron}-to-SZ flux ratio, as a function of frequency, is therefore 
\begin{equation}
\begin{split}
\dfrac{S_{\nu}^{\mathrm{sync.}}}{S_{\nu}^{\mathrm{SZ}}} & \approx ~ \mathcorrect{-9\times 10^4}~ 
	\left(\dfrac{\xi_{e/p}}{0.05}\right)~ \left(\dfrac{\mach}{3}\right)  
	\left(\dfrac{T_{\mathrm{u}}}{1~ \mathrm{keV}}\right)^{1/2} \left(\dfrac{\cal W}{100~ \mathrm{kpc}}\right)^{-1} \\
	& \times (1+z)^{-(4+\delta/2)}~ \dfrac{B_{\mathrm{relic}}^{1+\delta/2}}{B_{\mathrm{CMB}}^2 + B_{\mathrm{relic}}^2}  
	\left(\dfrac{\nu}{1.4 ~\mathrm{GHz}}\right)^{-(2+\delta/2)}.
\end{split}	
\label{eq:Fratio}
\end{equation}
The gas density and relic LLS dependence is absent from the flux ratio, its primary dependence is on the uncertain particle acceleration efficiency term and the e/p ratio, and to a lesser extent on the shock geometry and the relic magnetic field. 
With the simplifying assumption $\eta(\mach) \approx 3\times 10^{-4}\mach^3$, the Mach dependence has reduced to a simple linear scaling. The distances cancel out yielding $D_A^2/D_L^2 = (1+z)^{-4}$, with additional $(1+z)$ dependence from the \syn $K$-correction.

\subsection{Magnetic field from the \rev{synchrotron}-to-SZ flux ratio}
\label{sec:Bfield}

As one sees from Eq. \ref{eq:Fratio}, the flux ratio between the SZ and \rev{synchrotron} signals can be used to make an estimation for the relic magnetic field. This will require several assumptions which are as-yet untested, e.g. the CRe injection spectrum and the acceleration efficiency at low Mach numbers. However, we wish to emphasize that once these assumptions are made -- which can be 
 motivated based on current theory and observations -- then the SZ-to-\rev{synchrotron} flux ratio at GHz frequencies can be a particularly easy-to-measure tool for putting constraints on magnetic fields (or the shock acceleration efficiency).   
\rev{The direct observational method for constraining ICM magnetic fields, through the measurement of Faraday rotation angles of background polarized radio sources, have yielded results for only a handful of cases (e.g. \citealt{Murg04}, \citealt{Bona13}). Indirect methods like measuring the inverse-Compton emission in hard X-rays so far could only yield upper limits (e.g. \citealt{Bar15}). Connecting the thermal signature of the shock with its non-thermal \syn emission can provide another indirect yet independent method.}
%
%Currently there are only two direct observational methods for constraining the $B$-field in clusters: {\it (i)} through measurement of the Faraday rotation angles of background sources, which have yielded successful measurements for a handful of cases (e.g. \citealt{Murg04}, \citealt{Bona13}), and {\it (ii)} through measuring the inverse-Compton emission at hard X-rays, which so far could only yield upper limits (e.g. \citealt{Bar15}). Connecting the thermal signature of the shock with its non-thermal radio emission can provide a third, independent method. 

The parameter dependence in Eq. \ref{eq:Fratio} for the flux ratio can be made clearer if we take the strong field limit, $B_{\mathrm{relic}} \gg B_{\mathrm{CMB}}$, such that the term containing the magnetic field becomes approximately $B_{\mathrm{relic}}^{\delta/2 - 1}$. Then from Eq. \ref{eq:Fratio} we get for the magnetic field strength
\begin{equation}
B_{\mathrm{relic}} \propto \left[-\left(\dfrac{S_{\nu}^{\mathrm{sync.}}}{S_{\nu}^{\mathrm{SZ}}}\right)~
	\dfrac{{\cal W} ~ \nu_{\mathrm{obs}}^{2+\delta/2}}{\mach~ \xi_{e/p}~ T_{\mathrm{u}}^{1/2}}~\right]^{2/(\delta-2)} ~
	(\mathrm{for}~ B_{\mathrm{relic}} \gg B_{\mathrm{CMB}})
\label{eq:Bfield}
\end{equation}
The flux ratio term in parenthesis is just the inverse of contamination fraction: $C(\%) \equiv 100\times (S_{\nu}^{\mathrm{SZ}}/S_{\nu}^{\mathrm{sync.}})$, which is easy to determine observationally. With the current broadband receivers a single low frequency ($\lesssim 1$ GHz) measurement will provide the true \syn flux as well as the spectral slope, and hence the value of $\delta$, provided the low-frequency relic spectrum is close enough to a power-law. The Mach number is derived from the spectral slope as $\delta = 2(\mach^2+1)/(\mach^2-1)$, or alternatively, independent X-ray or SZ measurements can provide the value of $\mach$. 
\rev{The electron-to-proton ratio can be considered fixed at some universal value of roughly 5\%.} 
Then one single high-frequency measurement of the radio flux will yield the departure from the \syn power-law and hence the SZ-to-\rev{synchrotron} flux ratio. 

This procedure basically uses the SZ signal to constrain the pre-shock gas properties in a DSA-based \rev{synchrotron} model (Eq. \ref{eq:rmodel}). We emphasize that the same task can also be done with X-ray spectral imaging data, although obtaining upstream temperatures from X-ray data can be difficult, particularly at high redshifts. 
The temperature dependence in Eq. \ref{eq:Bfield} is weak, for massive clusters near 1 Mpc radius we generally have $T_{\mathrm{u}} \sim 2-3$ keV, so the uncertainty from the temperature term will be small. We have eliminated the density dependence, which can vary by a much wider margin than temperature at the cluster outskirts (see discussion in \citealt{Bona13} for the Coma relic). The relic width ${\cal W}$ is  determined readily from high-resolution radio observation.  Therefore, if the Mach dependence of particle acceleration efficiency is known from a model or can be constrained using independent observations, a constraint on the relic magnetic field will follow.

As a demonstration, we derive the magnetic field strength for the \sausage relic, from Eq. \ref{eq:Fratio}, to show that it  provides reasonable values despite the assumptions we have made. 
From our modeling of the SZ contamination in \sausage radio measurements (Section \ref{sec:sausage}), we get roughly 3\% contamination for $\mach = 3$ for an observation at 10 GHz, i.e., $S_{\nu}^{\mathrm{SZ}}/S_{\nu}^{\mathrm{sync.}} \approx -0.03$. 
The particle energy index is $\delta = 2.5$ for this choice of Mach number. 
The measurement by \citet{vanWe10} show a mean half-power width of the relic emission at 0.6 GHz roughly ${\cal W} \approx 70$ kpc, and we take the un-shocked (upstream) temperature $T_{\mathrm{u}} = 2$ keV. Then using our approximation for the acceleration efficiency $\eta(\mach)$ at $\mach \sim 3$, we get a value for the \sausage relic \rev{magnetic field strength of roughly $B_{\mathrm{relic}} \approx 2.5$ $\mu$G (or 3 $\mu$G for relic width ${\cal W}=100$ kpc). This is somewhat lower than the equipartition magnetic field at this cluster's redshift, and fully consistent with the best-fit estimate of \citet{vanWe10}.}

%%%%%%%%%%%%%%%%%%%%%%%%%%%%%%%%%%%%%%%%%%%%%%%%%%%%%%%%%%%%%%%%%%%%%%%%%%%%%%%%%%%%%%%
\section{Summary and conclusions}
\label{sec:con}
%%%%%%%%%%%%%%%%%%%%%%%%%%%%%%%%%%%%%%%%%%%%%%%%%%%%%%%%%%%%%%%%%%%%%%%%%%%%%%%%%%%%%%%

Motivated by some recent observations of galaxy cluster radio relics at frequencies above 10 GHz, we investigated how the Sunyaev-Zel'dovich (SZ) effect signal might affect these observations if the radio relics are manifestations of cluster merger shocks. This problem has not been addressed in detail so far, and we developed a semi-analytical model for quantitative prediction. The primary argument came from the fact that shocks at radio relics will boost the ambient pressure of the intracluster medium locally by roughly an order of magnitude (scaling as $\sim \mach^2$). Thus even though relics are located in the cluster outskirts where ambient pressure is low, the local shock boost will generate a negative SZ signal that can be significant at cm-wavelengths. Measurements made from interferometers will not be fully immune to the flux contamination, as the change in the SZ signal is localized roughly within the same angular scales as the \syn emission. 
The main steps for our modeling were as following:

\begin{itemize}
\item We assumed a spherical shock model for creating the pressure boost at the shock front, which is added on top of a cluster-wide spherical pressure that follows the universal GNFW profile, scaled to the cluster mass and redshift. These were then projected together to obtain the SZ decrement at the required frequency. \\

\item We modeled the radio \syn emission semi-analytically with a lognormal emissivity profile, that we matched to the published low-frequency radio data. The shape, length and the total power of the \syn emission were fitted on case-by-case basis for individual cluster predictions. In addition, we used a fiducial relic geometry, and an empirical scaling law between the total relic power and cluster mass, to compute flux contamination trends against redshift and mass. \\

\item We performed realistic simulations to mimic observations made by interferometers, and found that also interferometric imaging will report flux loss due to the SZ effect. Moreover, most interferometers will be affected by a lack of total power measurement due to the large angular sizes of radio relics. The actual interferometric measurements can be higher or lower than the `idealized' contamination predictions that we reported, depending on the array configuration, deconvolution methods etc, but will be roughy of similar magnitude.
%the reduction in flux due to the SZ effect will still be unavoidable.
\end{itemize}

\smallskip
Our results can be divided in two parts:  
first is a word of caution for the observers dealing with current and future radio relic data at cm-wavelengths, and  second is a proposal for future theoretical works to model the SZ-to-\rev{synchrotron} flux ratio in greater detail. 

\subsection*{Caution for observers}
\begin{itemize}
\item SZ flux contamination cannot be neglected for radio relic observations at $\nu_{\mathrm{obs}} \gtrsim 10$ GHz in almost all the cases. We compared the most recent data for the \sausage and \toothbrush relics up to 30 GHz, and found that SZ contamination can explain the observed spectral steepening to a large extent. At 10 GHz their flux contamination is roughly 10\%, comparable to the current measurement uncertainties, but at 30 GHz the contamination reaches 50\% level, i.e. only half of the original \syn flux is observed.  \\

\item We presented estimates for a set of double relics and found their contamination to scatter between $10-70$\% level. We also investigated two famous radio relics in the local universe, in the A2256 and Coma clusters. Their \syn profile modeling indicated that these single relics might be viewed at large angles, especially in A2256. The SZ flux contamination is also large: for A2256 it is roughly 30\%  at 10 GHz and close to 100\% at 20 GHz, whereas for Coma it is lower, about 6\% and 40\% at these frequencies. Such high contaminations can be a concern for single-dish observations of local universe radio relics.  \\

\item Due to the redshift independence of the SZ brightness, flux contamination goes up rapidly with redshift, and we found roughly an order of magnitude increase between redshifts 0 and 1. 
In addition, an empirical mass scaling of the relic \syn power suggested a higher contamination in low mass systems. We applied our model to the highest redshift radio relic known, the NW relic in the El Gordo cluster ($z=0.87$), and found it to have roughly 50\% flux contamination at 15 GHz, and its \syn flux would be completely eclipsed by the negative SZ signal near 30 GHz. This redshift trend will be particularly relevant for the large number of intermediate- to high-$z$ radio relics that are expected to be discovered by upcoming surveys with ASKAP, MeerKAT and SKA. 
\end{itemize}

Thus it might be premature to consider non-standard particle acceleration scenarios or other physical processes are at play at  high frequencies based solely on a direct measurement of spectral steepening for radio relics. There can be some degree of intrinsic steepening occurring still, e.g. as hinted from the \sausage or the El Gordo NW relic data. However, the amplitude required by the non-standard scenarios can only be constrained in a meaningful way once the impact of the SZ effect has been robustly constrained in conjunction with instrument-specific imaging techniques.

\subsection*{Challenge for theorists}
Far from being a simple ``nuisance'' for radio observations, the SZ decrement at cm-wavelengths offers probably the best evidence for the relic-shock connection. We outlined some possible advantages that might be gained from the measurement, and also a more precise modeling, of the \rev{synchrotron}-to-SZ flux ratio.

\begin{itemize}
\item If the spectral slope of the \syn spectrum at low frequencies can be robustly constrained, then the observed amplitude of the steepening can inform about the Mach number of the shock, in a way which is complementary to the standard radio estimate using the spectral slope from the \syn spectrum (Sec. \ref{sec:nontherm}). \\

\item We provided a simple analytical approximation for the \rev{synchrotron}-to-SZ flux ratio (Sec. \ref{sec:theory}). 
%that matches results from our more detailed semi-analytical modeling reasonably well. 
This step required connecting the number (and energy) of the cosmic-ray electrons to that of the thermal electrons present in the intracluster medium, and is theoretically uncertain. We assumed a DSA-based model for the \syn emission which includes several unknowns like the shock acceleration efficiency at low Mach numbers or the cosmic ray e/p ratio.  The resulting flux ratio depends weakly on the thermal gas temperature, is independent of its density, and can provide certain practical advantages for constraining theoretical models for particle acceleration. \\

\item We proposed that the \rev{synchrotron}-to-SZ flux ratio can be a tool for measuring the relic magnetic fields in the cluster outskirts, or the shock acceleration efficiency if magnetic field is known (Sec. \ref{sec:Bfield}). Even though making use of several restrictive theoretical assumptions for the analytical approximation, we were able to get realistic magnetic field strengths when plugging-in flux contamination values obtained in the  earlier part of our work. For the \sausage relic the estimate was roughly $3 ~\mu$G.  
Our suggestion is that through an accurate theoretical modeling of the \rev{synchrotron}-to-SZ flux ratio important advances can be made towards understanding of the radio relic origin, and the cm-wavelength window will be critical for testing these theoretical models. 
\end{itemize}

%%%%%%%%%%%%%%%%%%%%%%%%%%%%%%%%%%%%%%%%%%%%%%%%%%%%%%%%%%%%%%%%%%%%%%%%%%%%%%%%%%%%%%%
\smallskip
\begin{acknowledgements}
We thank Torsten En{\ss}lin, Marcus Br{\"u}ggen and Annalisa Bonafede for helpful discussions, Monica Trasatti for sharing unpublished Coma relic results, and Tracy Clarke for providing data for the A2256 relic. 
\rev{We thank the anonymous referee for a  thorough reading of the manuscript and pointing out a mistake in Eq. \ref{eq:Fratio}.} 
FV acknowledges support from the grant FOR1254 and VA876/3-1, JE support from the grant SFB956, and MS support from the Transregio Program TR33, all of the Deutsche Forschungsgemeinschaft (DFG). 
%This research has made use of NASA's Astrophysics Data System. 
\end{acknowledgements}

%%%%%%%%%%%%%%%%%%%%%%%%%%%%%%%%%%%%%%%%%%%%%%%%%%%%%%%%%%%%%%%%%%%%%%%%%%%%%%%%%%%%%%%
{\small
\bibliographystyle{aa}
\bibliography{szrelic_references}
}

\end{document}